\documentclass[letterpaper,twocolumn,prl,aps,superscriptaddress,amsmath,amssymb,floatfix]{revtex4}
\usepackage[latin9]{inputenc}
\usepackage{color}
\usepackage{amsmath}
\usepackage{amssymb}
\usepackage{amsthm}
\usepackage{mathrsfs}
\usepackage{graphicx}
\usepackage{esint}
\usepackage[unicode=true,
 bookmarks=true,bookmarksnumbered=false,bookmarksopen=false,
 breaklinks=false,pdfborder={0 0 1},backref=false,colorlinks=true]
 {hyperref}
\hypersetup{linkcolor=magenta,urlcolor=blue,citecolor=blue,pdfstartview={FitH},hyperfootnotes=false}

\makeatletter

\usepackage{textcomp}
\usepackage{epstopdf}
\usepackage{amsfonts}

\pdfpageheight\paperheight
\pdfpagewidth\paperwidth

\@ifundefined{textcolor}{}{%
 \definecolor{BLACK}{gray}{0}
 \definecolor{WHITE}{gray}{1}
 \definecolor{RED}{rgb}{1,0,0}
 \definecolor{GREEN}{rgb}{0,1,0}
 \definecolor{BLUE}{rgb}{0,0,1}
 \definecolor{CYAN}{cmyk}{1,0,0,0}
 \definecolor{MAGENTA}{cmyk}{0,1,0,0}
 \definecolor{YELLOW}{cmyk}{0,0,1,0}
}

\usepackage{xcolor}\usepackage{soul}
\newcommand{\ket}[1]{\ensuremath{\left|#1\right\rangle}}

\definecolor{blue}{rgb}{0,0,1}
\definecolor{red}{rgb}{1,0,0}
\definecolor{green}{rgb}{0,1,0}

\makeatother

\begin{document}
\title{High-efficiency arbitrary quantum operation on a high-dimensional quantum system}

\author{W.~Cai}
\thanks{These two authors contributed equally to this work.}
\affiliation{Center for Quantum Information, Institute for Interdisciplinary Information Sciences, Tsinghua University, Beijing 100084, China}

\author{J.~Han}
\thanks{These two authors contributed equally to this work.}
\affiliation{Center for Quantum Information, Institute for Interdisciplinary Information Sciences, Tsinghua University, Beijing 100084, China}

\author{L.~Hu}
\affiliation{Center for Quantum Information, Institute for Interdisciplinary Information Sciences, Tsinghua University, Beijing 100084, China}

\author{Y.~Ma}
\affiliation{Center for Quantum Information, Institute for Interdisciplinary Information Sciences, Tsinghua University, Beijing 100084, China}

\author{X.~Mu}
\affiliation{Center for Quantum Information, Institute for Interdisciplinary Information Sciences, Tsinghua University, Beijing 100084, China}

\author{W.~Wang}
\affiliation{Center for Quantum Information, Institute for Interdisciplinary Information Sciences, Tsinghua University, Beijing 100084, China}

\author{Y.~Xu}
\affiliation{Center for Quantum Information, Institute for Interdisciplinary Information Sciences, Tsinghua University, Beijing 100084, China}

\author{Z.~Hua}
\affiliation{Center for Quantum Information, Institute for Interdisciplinary Information Sciences, Tsinghua University, Beijing 100084, China}

\author{H.~Wang}
\affiliation{Center for Quantum Information, Institute for Interdisciplinary Information Sciences, Tsinghua University, Beijing 100084, China}

\author{Y.~P.~Song}
\affiliation{Center for Quantum Information, Institute for Interdisciplinary Information Sciences, Tsinghua University, Beijing 100084, China}

\author{J.-N.~Zhang}
\affiliation{Beijing Academy of Quantum Information Sciences, Beijing 100193, China}

\author{C.-L.~Zou}
\email{clzou321@ustc.edu.cn}
\affiliation{Key Laboratory of Quantum Information, CAS, University of Science and Technology of China, Hefei, Anhui 230026, China}

\author{L.~Sun}
\email{luyansun@tsinghua.edu.cn}
\affiliation{Center for Quantum Information, Institute for Interdisciplinary Information Sciences, Tsinghua University, Beijing 100084, China}

\begin{abstract}
The ability to manipulate quantum systems lies at the heart of the development of quantum technology. The ultimate goal of quantum control is to realize arbitrary quantum operations (AQuOs) for all possible open quantum system dynamics. However, the demanding extra physical resources impose great obstacles. Here, we experimentally demonstrate a universal approach of AQuO on a photonic qudit with minimum physical resource of a two-level ancilla and a $\log_{2}d$-scale circuit depth for a $d$-dimensional system. The AQuO is then applied in quantum trajectory simulation for quantum subspace stabilization and quantum Zeno dynamics, as well as incoherent manipulation and generalized measurements of the qudit. Therefore, the demonstrated AQuO for complete quantum control would play an indispensable role in quantum information science.

\end{abstract}


\maketitle


Behind the flourish of quantum technology~\cite{Nielsen,Google2019} is the mature quantum control technique~\cite{VanMeter2013,Matsuura2019}, which allows one to manipulate the quantum states of a physical system with unprecedented precision. The conquests in quantum domain proceed from arbitrary quantum state preparation~\cite{Law1996,Ben-Kish2003,Hofheinz2009} to arbitrary quantum gates~\cite{Khaneja2005,Dawson2005,Krastanov2015,Heeres2017}, but are mostly limited to closed quantum systems. However, isolated quantum systems do not exist in practice. On one hand, quantum systems are surrounded by the environment and unavoidably exposed to noise. On the other hand, the control and readout of the quantum systems are necessary for quantum tasks, requiring the communication with external systems. Therefore, practical quantum systems are open and their states should be described by density operators, and physical processes acting on the quantum systems are generally described by quantum operations~\cite{Nielsen} as $\mathcal{E}\left(\rho_{s}\right)=\sum_{j}E_{j}\rho_{s}E_{j}^{\dagger}$, where $\rho_{s}$ is the density matrix of the quantum system with $d$ dimensions and $E_{j}$ is the Kraus operator ($\sum_{j} E_{j}^{\dagger}E_{j}=I_{d\times d}$ with $I_{d\times d}$ being the $d$-dimensional identity matrix).


\begin{figure}
\centering{}\includegraphics[width=1\columnwidth]{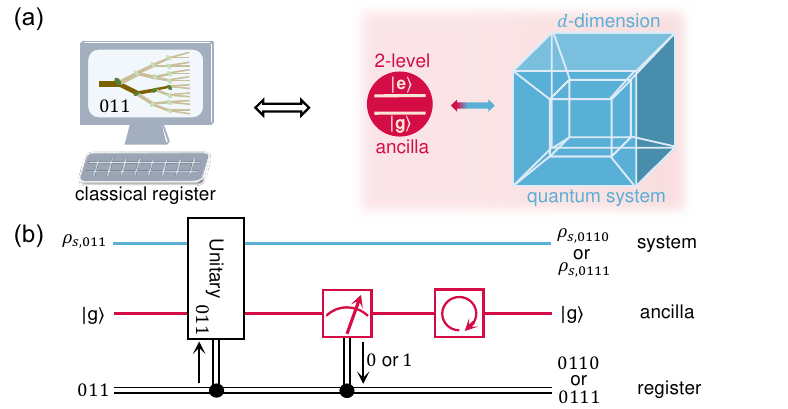} \caption{Scheme of arbitrary quantum operation (AQuO) on a high-dimensional
quantum system with minimum resource. (a) The architecture:
A $d$-dimensional quantum system is controlled by only one two-level
ancilla qubit and a classical register. (b) The elementary
quantum circuit for AQuO. A unitary gate is chosen according to the
digits in the classical register and acted on the composite of the quantum system and the ancilla. The ancilla is projectively measured with binary outcomes that are extracted by the register, and then the ancilla is reset to the ground state. The circuits can be implemented repetitively with the measurement results of the ancilla stored in the register for the next implementation of a new unitary
gate.}
\label{fig:concept}
\end{figure}

Great attention has already been drawn for realizing arbitrary quantum operations
(AQuOs) and universal quantum control of open quantum systems~\cite{lloyd1996universal,Lloyd2001PRA,Koch2016,Shen2017PRB}. Although direct control of the interaction between the quantum system and the environment to realize certain open quantum system dynamics has been demonstrated~\cite{Lu2017,Touzard2018PRX}, such an analog approach for AQuO is experimentally challenging due to the lack of the capability for arbitrary Hamiltonian engineering. Quantum operations of a $d$-dimensional system could also be realized digitally by unitary gates acting on a dilated Hilbert space. For example, rank-$2$ AQuOs are experimentally demonstrated in a trapped-ion system by using a two-dimensional ancilla, and it allows the dissipative quantum state preparation and the simulation of dynamical maps~\cite{Barreiro2011An,Schindler2013}. However, at least $m$ ancillary degrees of freedom are required for simulating the environment and realizing a rank-$m$ quantum operation~\cite{Nielsen} ($m\leq d^{2}$, see Ref.~\cite{Supplement}\nocite{DeFouquieres2011,Hatridge,Murch,Gilchrist2005,Petruccione2002,gardiner2004quantum,Wiseman2005,Bhandari2016,Franson2004,Johansson2012,Johansson2013,Bent2015PRX}), imposing tremendous resource overhead. Very recently, Shen $et~al.$ theoretically prove an efficient scheme to realize AQuOs with an arbitrary rank by adaptive control with only one two-dimensional ancilla and a $\log_{2}d$-scale circuit depth~\cite{Shen2017PRB}.

Here, we experimentally demonstrate AQuOs on a qudit with a dimension
$d=4$. With the assistance of only one ancilla qubit, we validate
the adaptive control scheme~\cite{Lloyd2001PRA,Shen2017PRB}
through high-fidelity universal unitary gates and real-time feedback control
in a superconducting quantum circuit. Based on the AQuOs,
we experimentally simulate the quantum trajectory of the qudit, and
show the important applications of the AQuOs in subspace stabilization and
quantum Zeno effect~\cite{Facchi2002,Bretheau2015}. Complete quantum control is illustrated by implementing a quantum task, including state preparation, incoherent quantum information processing, and detection through a rank-$16$ symmetric informationally complete (SIC) positive operator-valued measure (POVM)~\cite{Renes2004SICPOVM}. Our results on complete control of quantum systems could be easily generalized to other experimental platforms, such as trapped-ion~\cite{Fluhmann2019} and phononic~\cite{Chu2017} systems, and open up new possibilities in exploring quantum computation, quantum simulation, and quantum metrology~\cite{Pirandola2018,Alexeev2019,Altman2019}.

Figure~\ref{fig:concept}(a) sketches the architecture for
realizing the AQuO on an arbitrarily high dimensional quantum system, which
consists of a two-level ancilla qubit coupling with the system and a classical register that communicates with the ancilla.
The elementary quantum circuit of the AQuO is shown in
Fig.~\ref{fig:concept}(b) and only requires three types
of operations: a unitary gate $U$ on the composite of the quantum system and the ancilla (total dimension of $2d$), a projective measurement of the ancilla including an extraction of the outcome, and reset of the ancilla after the measurement. For each implementation of the elementary quantum circuit, we have the state of the composite after the unitary gate as $U\left(\rho_{s}\otimes\left|g\right\rangle \left\langle g\right|\right)U^{\dagger}=E_{0}\rho_{s}E_{0}^{\dagger}\otimes\left|g\right\rangle \left\langle g\right|+E_{1}\rho_{s}E_{1}^{\dagger}\otimes\left|e\right\rangle \left\langle e\right|+E_{0}\rho_{s}E_{1}^{\dagger}\otimes\left|g\right\rangle \left\langle e\right|+E_{1}\rho_{s}E_{0}^{\dagger}\otimes\left|e\right\rangle \left\langle g\right|$. Here, $\left|g\right\rangle $ and $\left|e\right\rangle $ are the ground and the excited states of the ancilla, respectively, and $E_{j}$ is the Kraus operator according to the ancilla measurement outcome ($j\in\left\{ 0,1\right\}$ corresponding to \{\ket{g},\ket{e}\}). By choosing an appropriate $U$ and tracing out the ancilla regardless the outcome of the measurement, a rank-$2$ AQuO
\begin{equation}
\mathcal{E}\left(\rho_s\right)=E_{0}\rho_s E_{0}^{\dagger}+E_{1}\rho_s E_{1}^{\dagger}\label{eq:Rank2}
\end{equation}
can be realized. From the aspect of open quantum system dynamics,
the ancilla plays two roles in the AQuO. First, the ancilla can be treated as a quantum dice, i.e., a quantum random number generator
for mimicking noise and inducing stochastic quantum jumps of the system~\cite{Wiseman2014Quantum}. Second, by monitoring
the quantum system through the ancilla, the entropy of the system
could be dumped through the ancilla into the classical register. For
example, a quantum error correction operation could restore the pure quantum
state from a mixed state~\cite{Nielsen}.

To experimentally demonstrate the AQuOs of a high dimensional quantum
system, we carry out the architecture in Fig.~\ref{fig:concept}(a)
via a superconducting quantum circuit, which consists of a transmon
qubit and a high-quality microwave cavity~\cite{Paik2011,Devoret2013,Hu2019,Ma2020}. The cavity
provides a photonic qudit by exploring its infinitely large Hilbert
space of Fock states, and the AQuO on the qudit is realized through
the transmon qubit serving as the ancilla and a field programmable gate array (FPGA) serving as the classical register~\cite{Hu2019}. The FPGA not only records the measurement outcomes of the ancilla, but also executes classical
logic in real time and adaptively sends appropriate control
pulses to the composite cavity-transmon system for arbitrary unitary
gates~\cite{Khaneja2005,Heeres2017}. All required gates
on the system could be implemented with high fidelities and negligible
latency, thus allowing the repeating and cascading of the quantum circuits (see
Ref.~\cite{Supplement}).

\begin{figure*}
\centering{}\includegraphics[width=2\columnwidth]{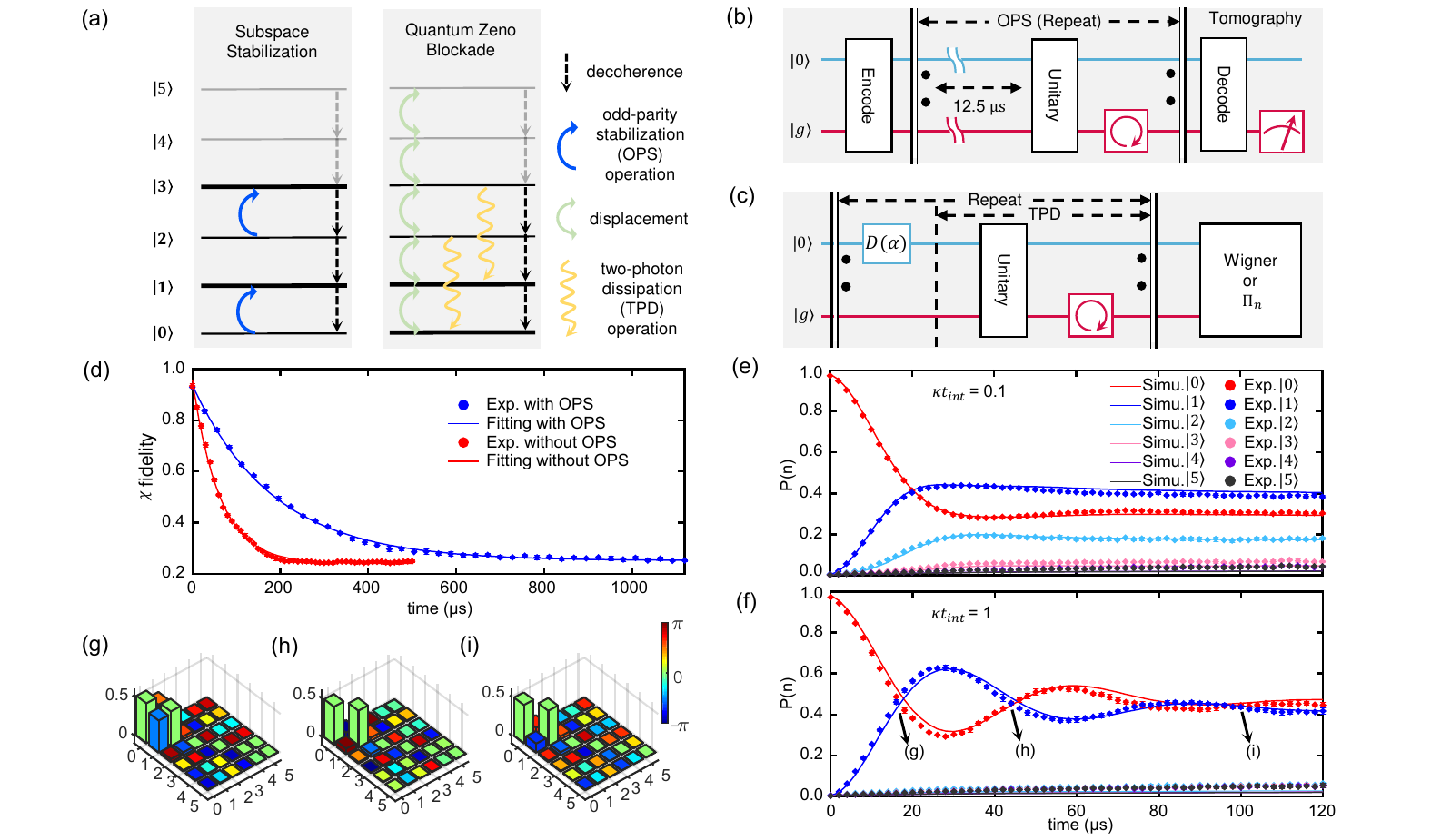} \caption{Experimental quantum trajectory simulation through AQuO. (a) The energy-level transition diagrams of rank-2 AQuOs, with the left and the right panels representing two distinct schemes for the odd-parity
subspace stabilization (b, d) and quantum Zeno blockade (c, e-i), respectively. (b) Experimental sequence for the subspace stabilization by  odd-parity stabilization (OPS) operations. (c) Experimental sequence for the quantum Zeno blockade, consisting of two-photon dissipation (TPD) and displacement $(D(\alpha))$ operations and Wigner tomography or photon-number measurement $\Pi_n$. (d) Results of the $\chi$ matrix fidelity decay in the odd-parity subspace. The process fidelity decay times $T_{1}$ = 184.3~$\mu$s and 61.9~$\mu$s for the system with and without repetitive OPS operations, respectively. (e-f) Time evolutions of Fock state populations $(\ket{0}$ to $\ket{5})$ for effective TPD rates of $\kappa/2\pi=8$~kHz (e) or $\kappa/2\pi=80$~kHz (f), a displacement value $\alpha=-0.1i$, and a repetition interval $t_{\rm{int}}=2~\mathrm{\mu s}$. Dots and lines present the experimental and the numerical results, respectively. (g-i) Density matrices reconstructed by Wigner tomography of the system at evolution times of 16~$\mu$s, 44~$\mu$s, and 100~$\mu$s in (f), respectively.}
\label{fig:continuous}
\end{figure*}

\begin{figure*}
\centering{}\includegraphics[width=2\columnwidth]{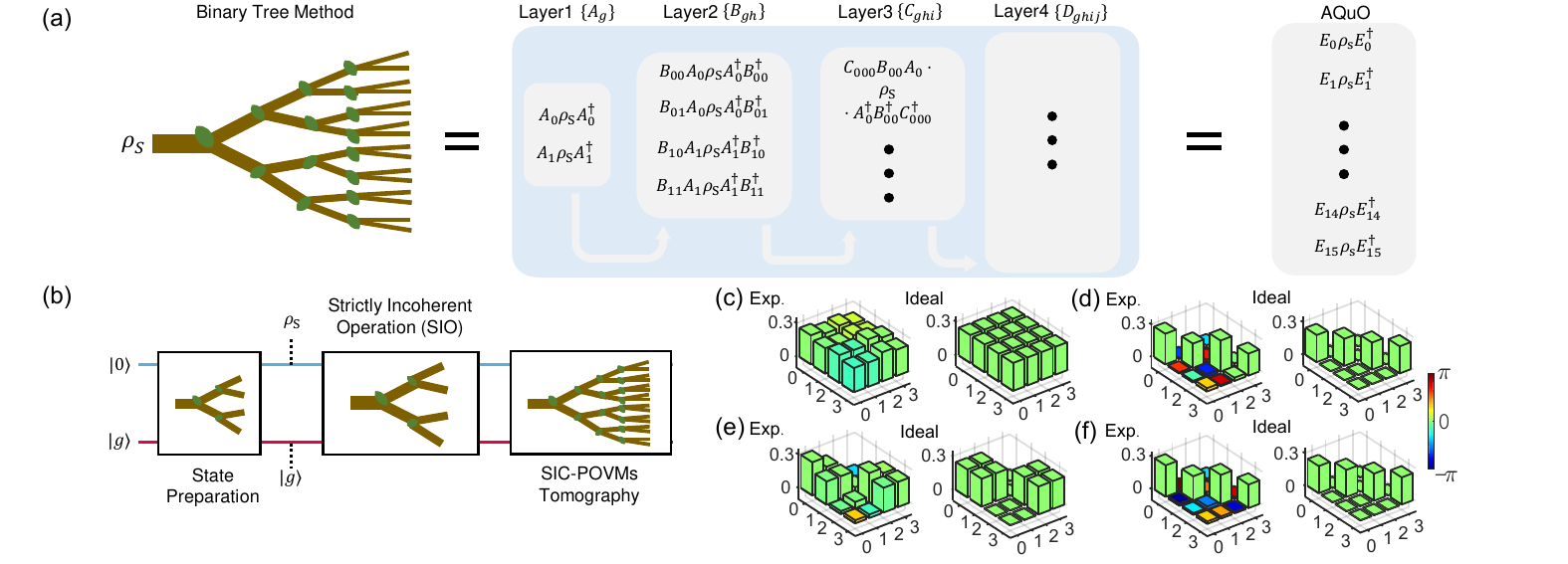} \caption{AQuO with adaptive control of the ancilla. (a) The binary-tree illustration of the full-rank AQuO construction for a photonic qudit with $d=4$. Each leaf represents an implementation of a rank-2 AQuO with a given unitary gate on the composite of the quantum system and the ancilla, and the evolution of the system follows one of the two branches according to the binary measurement outcomes of the ancilla (0 or 1). For a four-layer binary tree,
there are four measurements that produce $2^{4}=16$ possible outcomes (0000
to 1111), and eventually up to 16 Kraus operators could be
realized by $2^{4}-1$ unitaries (leaves). (b) An example quantum protocol for the manipulation of quantum coherence, which consists
of arbitrary state preparation, strictly incoherent operation
(SIO), and output state tomography through a symmetric informationally
complete positive operator-valued measure (SIC-POVM). (c) and (d) Density matrices of the photonic qudit with fidelities of $92.5\%$ and $98.8\%$ for the maximum coherent state $\ket{\psi_{\rm{mcs}}}$ and the maximally mixed state $\rho_{\rm{mms}}$, respectively. (e) and (f) Density matrices of the photonic qudit after implementing two SIOs (rank-2 and rank-4) on $\ket{\psi_{\rm{mcs}}}$, with the output state fidelity of the photonic qudit being $92.2\%$ and $99.1\%$, respectively.}
\label{fig:SIC}
\end{figure*}

We first test the rank-$2$ AQuOs in the spirit of stochastic
quantum trajectory simulation~\cite{Wiseman2014Quantum}. Under the continuous
monitoring by a Markovian environment, a quantum system evolves stochastically
conditional on the projection outcome of the environment, which has been widely
studied in theory by the Monte-Carlo method~\cite{Wiseman2014Quantum}. Interpreting
the ancilla as a monitor to the system and a quantum dice, we can simulate the stochastic quantum trajectories of the qudit in experiment via AQuOs. By repetitively erasing the ancilla, the ancilla and the register together act as an infinitely large Markovian environment to the system without information backflow. In such a manner, we experimentally simulate two quantum trajectories on a photonic qudit ($d=4$): an odd-parity subspace stabilization and quantum Zeno blockade (QZB) due to two-photon dissipation. The corresponding quantum operations on the truncated space of Fock states are depicted in Fig.~\ref{fig:continuous}(a).

To stabilize the odd-parity subspace of the photonic qudit, we engineer proper
unitary gates to achieve the two Kraus operators: $E_{0}$
reserves the odd-parity subspace span $\{\left|1\right\rangle ,\left|3\right\rangle \}$ and $E_{1}$ converts the even-parity subspace to the odd-parity subspace
$\left\{ \left|0\right\rangle \rightarrow\left|1\right\rangle ,\left|2\right\rangle \rightarrow\left|3\right\rangle \right\} $. Due to the spontaneous decay of the cavity, single-photon losses of the photonic qudit will convert quantum states in the odd-parity subspace to the even-parity subspace, and eventually destroy the quantum information stored in the qudit. By repetitively implementing the odd-parity subspace stabilization operations (Fig.~\ref{fig:continuous}(b)), we expect that any state trajectories in the odd-parity subspace would have a longer lifetime. To compromise with the operation error in practice, which is mainly induced by the ancilla's decoherence, we choose an optimal interval time (12.5~$\mathrm{\mu s}$) between two adjacent AQuOs. To characterize the performance of the stabilization operation, we carry out process tomography in the odd-parity subspace averaged over an ensemble of trajectories and evaluate the $\chi$ fidelity by comparing with an ideal qubit encoded in $\{\ket{1},\ket{3}\}$. Figure~\ref{fig:continuous}(d) shows the exponentially decaying $\chi$ fidelities that demonstrate the coherence time of the quantum state in the odd-parity subspace is boosted by approximately 3 times with the stabilization operations (184.3~$\mathrm{\mu s}$), compared to that without the operations (61.9~$\mathrm{\mu s}$).

Significantly different dynamics are unraveled with a two-photon dissipation
operation realized by engineering two Kraus operators (see Ref.~\cite{Supplement}) that effectively induce jumps of Fock states $\left|n\right\rangle \rightarrow\left|n-2\right\rangle $ for $n\geq2$ (Fig.~\ref{fig:continuous}(a)). A continuous
implementation of such a quantum operation induces quantum Zeno effect~\cite{Facchi2002,Bretheau2015} that blockades the transitions from energy levels $\left\{ \left|0\right\rangle ,\left|1\right\rangle \right\}$
to other higher levels and thus realizes an equivalent two-level system in a harmonic oscillator. The QZB is revealed when exciting the
cavity with coherent drives through a displacement gate on the cavity. By alternatively implementing the engineered quantum operation and the displacement gate with a repetition interval of $t_{\rm{int}}=2~\mathrm{\mu s}$ (Fig.~\ref{fig:continuous}(c)), the evolution of the qudit shows the Rabi oscillation dynamics that mimics a two-level atom under a coherent drive, as expected for QZB (Fig.~\ref{fig:continuous}(f)). The populations of the qudit are confined in the lowest two levels, whose combined population is $0.942$ at the evolution time $t=20~\,\mu$s when $\kappa t_{\rm{int}}=1$. For a smaller $\kappa$ ($\kappa t_{\rm{int}}=0.1$), the QZB is weaker with a higher population leakage to other states, as shown in Fig.~\ref{fig:continuous}(e). The detailed density matrices are shown in Figs.~\ref{fig:continuous}(g-i). The QZB protocol can be extended to a larger Zeno subspace with Kraus operators that induce multi-photon jumps~\cite{PRResearch2020Patsch}. Non-Markovian environment could also be realized with the memory effect of an ancilla by partially resetting the ancilla to the ground state or selecting the unitary gate depending on the previous ancilla measurement outcomes~\cite{PRResearch2020Patsch}.

The extension of the AQuO to rank-$m$
($m\leq d^{2}$) could be implemented by introducing a $m$-dimensional
ancilla. However, this approach demands large physical resource overhead. Instead, as proposed in
Ref.~\cite{Shen2017PRB}, a rank-$m$ AQuO could be realized by an $n$-step
($n=\left\lceil \log_{2}m\right\rceil $) cascading of the elementary
quantum circuits in Fig.~\ref{fig:concept}(b) via
adaptive control and a recycling strategy of a two-level
ancilla. Regardless of the measurement outcomes of the ancilla, the achieved quantum operation is
\begin{equation}
\mathcal{E}\left(\rho_s\right)=\sum_{r_{1},r_{2},...,r_{n}}E_{n,r_{n}}^{\left(r_{n-1}...r_{1}\right)}...E_{1,r_{1}}\rho_s E_{1,r_{1}}^{\dagger}...E_{n,r_{n}}^{\dagger \left(r_{n-1}...r_{1}\right)},\label{eq:rank-m}
\end{equation}
where $E_{k,r_{k}}^{\left(r_{k-1}...r_{1}\right)}$ denotes the Kraus
operator due to the unitary $U_{k}^{\left(r_{k-1}...r_{1}\right)}$
at the $k$-th step according to the outcomes of all previous steps $r_{k-1}...r_{1}$ with binaries $r_{k}\in\left\{ 0,1\right\}$. Such an architecture is universal for arbitrarily high dimensional quantum
systems, and greatly saves the hardware overhead.

Figure~\ref{fig:SIC}(a) provides a binary-tree illustration of the scheme, where the system evolves along the branches in the $n$-layer binary tree according to the ancilla outcomes $r_{n}...r_{1}$ and each
bifurcation (leaf) represents an elementary rank-$2$ AQuO. So, there are $n$ measurements of the ancilla in sequence, which require $2^{n}-1$
unitary gates and produce $2^{n}$ outcomes. For the qudit with
$d=4$ considered in this work, $n\equiv2\log_{2}d=4$ layers are enough for
full-rank ($m=d^{2}=16$) AQuOs. All $d^{2}$ Kraus operators
$E_{k}=D_{ghij}C_{ghi}B_{gh}A_{g}$ can be realized through appropriate choice
of operators at each step $\left\{ A_{g},\,B_{gh},\,C_{ghi},\,D_{ghij}\right\}$
(see Ref.~\cite{Supplement}), where $g,h,i,j$ are the outcomes and $k=\left(ghij\right)_{2}$ in the binary representation.

In addition to the arbitrary state preparation and manipulation of
quantum information through Eq.~\ref{eq:rank-m}, the AQuO can also be translated to POVM $M_{k}=E_{k}^{\dagger}E_{k}$ when recording
the outcome $k$ with a probability $p_{k}=\mathrm{Tr}\left[E_{k}\rho_{s}E_{k}^{\dagger}\right]$~\cite{Nielsen,Shen2017PRB,Andersson2008}.
We next implement a quantum information processing task as an example
to illustrate the application allowed by the powerful tool of AQuOs.

Figure~\ref{fig:SIC}(b) depicts the protocol for manipulating the quantum coherence of a qudit. The experimental
procedure includes the state preparation, strictly incoherent operations
(SIO)~\cite{Streltsov2017}, and a SIC-POVM~\cite{Renes2004SICPOVM} for output state tomography, and all steps are realized through AQuOs. The maximum coherent state $\ket{\psi_{\rm{mcs}}} =\frac{1}{2}(\ket{0}+\ket{1}+\ket{2}+\ket{3})$
and the maximally mixed state $\rho_{\rm{mms}}=\frac{1}{4}I_{4\times4}$ are chosen as examples and prepared with fidelities  $\mathcal{F}=92.5\%$ and $98.8\%$, respectively, as shown in Figs.~\ref{fig:SIC}(c) and \ref{fig:SIC}(d). The quantum coherences of the two experimental states are quantified by the relative entropy of coherence~\cite{Streltsov2017}  $\mathcal{C}=1.754$ and $0.018$, respectively. The coherence of the system can be manipulated by the AQuO. Here we demonstrate two SIOs with ranks of $2$ and $4$, respectively (see Ref.~\cite{Supplement}), which could also be used to measure the generalized parity of a bosonic mode. The performances of the two SIOs are tested by evaluating the fidelity and quantum coherence for the output of $\ket{\psi_{\rm{mcs}}}$. Here, the SIC-POVM for $d=4$ dimensional system contains $16$ elements, corresponding to a rank-16 quantum operation, and thus is realized by a four-layer binary tree (Fig.~\ref{fig:SIC}(a)). More details about the SIC-POVM construction can be found in Ref.~\cite{Supplement}.

The results of the quantum coherence manipulation are shown in
Figs.~\ref{fig:SIC}(e) and \ref{fig:SIC}(f). For the rank-$2$ SIO, the output state $\mathcal{F}=92.2\%$ and $\mathcal{C}=0.694$, while for
the rank-$4$ SIO, the output state $\mathcal{F}=99.1\%$ and $\mathcal{C}=0.016$, proving a better coherence erasing ability of the rank-$4$ SIO. Note that the state tomography of the photonic qudit through a SIC-POVM can be completed in several minutes, in sharp contrast to several hours required by the conventional Wigner tomography method in our experiment (see Ref.~\cite{Supplement}), manifesting the advantage of AQuOs in practical quantum tasks.


Our AQuO can be extended to the case of coupling one control qubit to two manipulated cavities simultaneously, which may be of importance to realize dissipative evolution for generating and stabilizing various non-trivial two-body~\cite{PRA2013Reiter,PRA2013Leghtas,shankar2013autonomously,lin2013dissipative,PRX2016Liu,horn2018quantum,kumar2020engineering} and many-body states~\cite{Kraus2008Preparation,weimer2010rydberg,diehl2011topology,Roy2020Measurement}. The AQuO can be easily implemented in other spin-oscillator systems,
such as the trapped-ion system~\cite{Fluhmann2019} and the hybrid superconducting-phononic system~\cite{Chu2017}. Importantly, the AQuO provides a unified framework of open quantum system control. For examples, the recently demonstrated quantum error corrections that rely on the ancilla-induced Markovian dissipation for the correction~\cite{campagneibarcq2020,Royer2020stab,de2020error} can be considered as rank-two AQuOs and the repeated adaptive phase estimations to create the GKP state~\cite{Terhal2016} resemble higher-rank AQuOs. This work therefore presents a significant conceptual advance in quantum technology that is beneficial for quantum information processing~\cite{Nielsen,Alexeev2019}, quantum simulation~\cite{Hu2018channel,McArdle2020}, and quantum precision measurement~\cite{Pirandola2018}.

This work was supported by National Key Research and Development Program of China (Grants No. 2017YFA0304303 and No. 2017YFA0304504), Key-Area Research and Development Program of Guangdong Province (Grant No. 2020B0303030001), the National Natural Science Foundation of China (Grants No. 11925404, No. 11874235, No. 11874342, No. 11922411, and No. 12061131011), Anhui Initiative in Quantum Information Technologies (AHY130200), and Grant No. 2019GQG1024 from the Institute for Guo Qiang, Tsinghua University.

\end{document}


\clearpage\newpage{}

\newpage{}

\onecolumngrid
\renewcommand{\thefigure}{S\arabic{figure}}
\setcounter{figure}{0}
\renewcommand{\thepage}{S\arabic{page}}
\setcounter{page}{1}
\renewcommand{\theequation}{S\arabic{equation}}
\setcounter{equation}{0}
\setcounter{section}{0}
\title{Supplementary Materials for ``High-efficiency arbitrary quantum operation on a high-dimensional quantum system"}
\author{W.~Cai}
\thanks{These two authors contributed equally to this work.}
\affiliation{Center for Quantum Information, Institute for Interdisciplinary Information
Sciences, Tsinghua University, Beijing 100084, China}
\author{J.~Han}
\thanks{These two authors contributed equally to this work.}
\affiliation{Center for Quantum Information, Institute for Interdisciplinary Information
Sciences, Tsinghua University, Beijing 100084, China}
\author{L.~Hu}
\affiliation{Center for Quantum Information, Institute for Interdisciplinary Information
Sciences, Tsinghua University, Beijing 100084, China}
\author{Y.~Ma}
\affiliation{Center for Quantum Information, Institute for Interdisciplinary Information
Sciences, Tsinghua University, Beijing 100084, China}
\author{X.~Mu}
\affiliation{Center for Quantum Information, Institute for Interdisciplinary Information
Sciences, Tsinghua University, Beijing 100084, China}
\author{W.~Wang}
\affiliation{Center for Quantum Information, Institute for Interdisciplinary Information
Sciences, Tsinghua University, Beijing 100084, China}
\author{Y.~Xu}
\affiliation{Center for Quantum Information, Institute for Interdisciplinary Information
Sciences, Tsinghua University, Beijing 100084, China}
\author{Z.~Hua}
\affiliation{Center for Quantum Information, Institute for Interdisciplinary Information
Sciences, Tsinghua University, Beijing 100084, China}
\author{H.~Wang}
\affiliation{Center for Quantum Information, Institute for Interdisciplinary Information
Sciences, Tsinghua University, Beijing 100084, China}
\author{Y.P.~Song}
\affiliation{Center for Quantum Information, Institute for Interdisciplinary Information
Sciences, Tsinghua University, Beijing 100084, China}
\author{J.-N.~Zhang}
\affiliation{Beijing Academy of Quantum Information Sciences, Beijing 100193, China}
\author{C.-L.~Zou}
\email{clzou321@ustc.edu.cn}

\affiliation{Key Laboratory of Quantum Information, CAS, University of Science
and Technology of China, Hefei, Anhui 230026, China}
\author{L.~Sun}
\email{luyansun@tsinghua.edu.cn}

\affiliation{Center for Quantum Information, Institute for Interdisciplinary Information
Sciences, Tsinghua University, Beijing 100084, China}

\maketitle
\tableofcontents{}

\section{Experimental Device and Setup}

\noindent Our experimental system consists of a superconducting transmon
qubit dispersively coupled to two rectangular microwave cavities,
which are machined with high purity 5N5 aluminum. The transmon qubit
is used as the ancilla, fabricated on a sapphire substrate with an energy relaxation time of $30~\mathrm{\mu s}$ and a pure dephasing time of $120~\mathrm{\mu s}$. The storage cavity has a single-photon lifetime of $143\,\mathrm{\mu s}$ with its lowest four Fock levels being employed as the qudit with a dimension of $d=4$. The ancilla qubit can be read out with a high
fidelity through another cavity with a lifetime of 44~ns. The detailed geometry of the device and measurement setup can be found in our previous work~\cite{Hu2019}.

The system Hamiltonian is given by
\begin{equation}
H/\hbar=\omega_{\mathrm{s}}a^{\dagger}a+\omega_{\mathrm{a}}\ket{e}\bra{e}-\chi_{\mathrm{s}}a^{\dagger}a\ket{e}\bra{e},\label{eq:Hamiltonian}
\end{equation}
where $a$ and $a^{\dagger}$ are the ladder operators of the storage
cavity with a mode frequency $\omega_{s}/2\pi=7.634~\mathrm{GHz}$,
$\omega_{a}/2\pi=5.692~\mathrm{GHz}$ is the transition
frequency of the transmon between the ground state $\left|g\right\rangle $
and the excited state $\left|e\right\rangle $, and $\chi_{s}/2\pi=1.90~\mathrm{MHz}$
is the dispersive coupling strength between the transmon and the storage
cavity.

The general architecture and scheme for realizing the arbitrary quantum operation
(AQuO) are illustrated in Fig.~1 in the main text. The AQuO can be realized in three steps: (i) a unitary gate on the composite of the quantum system and the ancilla, (ii) the readout of the ancilla, and (iii) the reset of the ancilla.

Step (i) corresponds to a universal control of the closed composite of the quantum system and the ancilla. According to the nonlinear interaction
Eq.~(\ref{eq:Hamiltonian}), an arbitrary unitary gate could be realized
by using selective number-dependent arbitrary phase gates (SNAP)~\cite{Krastanov2015,Heeres2017} on the ancilla in combination with displacement gates on the cavity. In this work, we instead use the gradient ascent pulse engineering (GRAPE) method~\cite{Khaneja2005,DeFouquieres2011} to numerically optimize the control pulses based on carefully calibrated experimental parameters to realize the unitary gates. We note that the GRAPE pulse length is mainly limited by the dispersive coupling strength $\chi_{s}$, and for a moderate $d\leq5$, the pulse length is pretty independent of the number $d$ of the Fock levels that are employed. However, when $d$ gets even larger, the higher-order terms of the cavity-transmon interaction and decoherence effects will start to matter, demanding higher control precision and longer pulse length. The control microwave drives are generated by field-programmable gate arrays (FPGAs) based on all previously recorded ancilla measurement outcomes in the classical register. The fast real-time adaptive control is the key technology for AQuO and is also realized with the FPGAs with home-made logics.



For step (ii), the quantum state of the ancilla qubit is read out with the
assistance of a readout cavity mode and a Josephson parametric amplifier~\cite{Hatridge,Murch}.
The readout results are requisite, registered, and processed by the FPGAs. With our current experimental apparatus, the readout fidelities are $>$99.9\% for $\ket{g}$ and 98.9\% for $\ket{e}$ of the ancilla qubit with a measurement time of $320~\mathrm{ns}$.

In step (iii), to reset the ancilla qubit requires no effect on the
photonic system during the resetting process. In our experiment, we use the measurement-based adaptive control to reset the ancilla qubit by applying a $\pi$-pulse conditional on the measurement outcome $\ket{e}$ of the ancilla.

In addition to the realization of AQuOs with the circuit in Fig.~1
in the main text, we also need to characterize the performance of
AQuOs by process tomography~\cite{Nielsen}. The process tomography of the AQuOs is realized with the assistance of the ancilla by preparing the qudit into a set of linearly independent basis states and performing the state tomography on the output of the AQuOs. The initialization and measurement of the qudit are achieved via the encoding and decoding gates based on the universal control of the composite photonic qudit-ancilla system via the GRAPE method. For example, the encoding and decoding gates for the odd-parity subspace stabilization are
\begin{equation}
\begin{split}\ket{0}\ket{g}\to\ket{1}\ket{g},\\
\ket{0}\ket{e}\to\ket{3}\ket{g},
\end{split}
\label{eq:S2}
\end{equation}
and
\begin{equation}
\begin{split}\ket{1}\ket{g}\to\ket{0}\ket{g},\\
\ket{3}\ket{g}\to\ket{0}\ket{e},
\end{split}
\label{eq:S3}
\end{equation}
respectively.

\section{Basics about quantum operations}

\subsection{Kraus operators}

The open quantum system dynamics has many representations, such as
superoperators in Liouville space, Kraus operators, Choi matrix, and
system-environment representations~\cite{Gilchrist2005,Nielsen}.
The Kraus representation of a quantum operation can be written as~\cite{Nielsen}
\begin{equation}
\rho_s\mapsto\mathcal{E}\left(\rho_s\right)=\sum_{j=0}^{l-1}E_{j}\rho_s E_{j}^{\dagger},\label{eq:SI-channel}
\end{equation}
where $1\le l\le d^{2}$ ($d$ is the Hilbert space dimension), $\rho_s$ is the density matrix of the quantum system, and the Kraus operators satisfy
\begin{equation}
\sum_{j=0}^{l-1}E_{j}^{\dagger}E_{j}=\mathbf{I},
\end{equation}
where $\mathbf{I}$ is the unity matrix.

Without being limited by the experimental resources, the conventional
approach to realize an AQuO on a $d$-dimensional quantum system is to
introduce an extra ancillary system serving as the environment to dilate the whole Hilbert space. A unitary gate on the composite quantum system-ancilla is then implemented followed by the ancilla reset. For an ancilla with a dimension
of $l$, the unitary gate is given by
\begin{equation}
U=\begin{pmatrix}E_{0} & * & ... & *\\
E_{1} & * & ... & *\\
... & * & ... & *\\
E_{l-1} & * & ... & *
\end{pmatrix},
\end{equation}
with the dimension of the whole system being $d\times l$. Here, `{*}'
means the specific matrix element is trivial because it will not affect
the target quantum system. We then have:
\begin{equation}
U\left[\ket{0}\bra{0}\otimes\rho_{s}\right]U^{\dagger}=\Sigma_{j_{1}j_{2}}\ket{j_{1}}\bra{j_{2}}\otimes E_{j_{1}}\rho_{s}E_{j_{2}}^{\dagger}.\label{eq:Kraus-Environment}
\end{equation}
Here, $E_{j}$ is the Kraus operator with a dimension of $d$, the subscript $j$ denotes the outcome of the ancilla in its orthonormal basis
$\ket{j}$ ($j\in\left\{ 0,...,l-1\right\} $) after implementing
the unitary gate $U$, and the initial state of the ancilla is $\left|0\right\rangle$. Tracing out the ancilla, one gets the quantum operation
\begin{equation}
\mathcal{E}(\rho_{s})=\Sigma_{j=0}^{l-1}E_{j}\rho_{s}E_{j}^{\dagger}.
\end{equation}

Note that the quantum operation is invariant with respect to the choice
of the ancilla basis. Therefore, for an arbitrary ancilla
basis $\left|\tilde{\psi}_{k}\right\rangle =\sum_{j=0}^{l-1}u_{kj}^{*}\left|j\right\rangle $
with an $\tilde{l}\times\tilde{l}$ unitary transformation $\boldsymbol{u}$
with a dimension $\tilde{l}\geq l$, we have the corresponding
Kraus operator representation as
\begin{align}
 & \mathrm{Tr}\left[U\ket{0}\bra{0}\otimes\rho_{\mathrm{s}}U^{\dagger}\right]\nonumber \\
 & =\sum_{k=0}^{\tilde{l}-1}\left\langle \tilde{\psi}_{k}\right|\Sigma_{j_{1}j_{2}}\ket{j_{1}}\bra{j_{2}}\otimes E_{j_{1}}\rho_{\mathrm{s}}E_{j_{2}}^{\dagger}\left|\tilde{\psi}_{k}\right\rangle \nonumber \\
 & =\sum_{k=0}^{\tilde{l}-1}\left[\sum_{j_{1}}\langle \tilde{\psi}_{k}\left|j_{1}\right\rangle E_{j_{1}}\right]\rho_{\mathrm{s}}\left\{ \left[\sum_{j_{2}}\langle \tilde{\psi}_{k}\left|j_{2}\right\rangle E_{j_{2}}\right]\right\} ^{\dagger}\nonumber \\
 & =\sum_{k=0}^{\tilde{l}-1}F_{k}\rho_{s}F_{k}^{\dagger},\label{eq:Kraus-transformation}
\end{align}
with $F_{k}=\sum_{j}u_{kj}E_{j}$.

This indicates that both the Kraus representation of an operation and the approach to realize a given quantum operation are not unique. The Kraus rank of an operation ($m$) is the number of the irreducible Kraus operator elements, which satisfies $1\leq m\leq d^{2}$.
Therefore, for AQuOs this conventional method needs an ancilla with
a dimension of $d^{2}$ and an arbitrary unitary gate on a $d^{3}$-dimensional
Hilbert space.

\subsection{Master equation}

It is well-known that the continuous evolution of an open quantum system
follows the master equation, in the Lindblad form as~\cite{Petruccione2002,gardiner2004quantum}
\begin{align}
\frac{d}{dt}\rho_s\left(t\right) & =\sum_{j=0}^{l-2}\kappa_{j}\left(2o_{j}\rho_s\left(t\right)o_{j}^{\dagger}-o_{j}^{\dagger}o_{j}\rho_s\left(t\right)-\rho_s\left(t\right)o_{j}^{\dagger}o_{j}\right).\label{eq:SI-masterEq-2}
\end{align}
Here, $\kappa_{j}$ and $o_{j}$ are the rate and the operator for the $j$-th decoherence process, respectively. To mimic an arbitrary decoherence
process of a quantum system, we could construct $l-1$ independent
environments $\sum_{\omega}b_{j}\left(\omega\right)$, with $b_{j}\left(\omega\right)$ denoting the bosonic operator for a continuum mode at a frequency $\omega$ of the $j$-th environment, and realize the system-environment coupling Hamiltonian $H_{j}=\sum_{\omega}\left[g_{j}\left(\omega\right)b_{j}\left(\omega\right)o_{j}^{\dagger}+g_{j}^{*}(\omega)b_{j}^{\dagger}\left(\omega\right)o_{j}\right]$.
For a Markovian environment, $g_{j}\left(\omega\right)\approx g_{j}$
is independent of $\omega$ and we could derive the Lindblad form
of the master equation by tracing over the environments, and the corresponding
decoherence rates $\kappa_{j}=g_{j}^{2}\rho_{j}$ with $\rho_{j}$
denoting the density of the continuum states for the $j$-th environment. Therefore,
such an analog approach to realize the open quantum system evolution
is resource demanding because of the challenges in the construction
of Markovian environments and arbitrary system-environment interaction
Hamiltonians.

\subsection{Master equation to AQuO conversion}

From another perspective, by discretizing the evolution into steps with
$dt\rightarrow0$, we have the dynamics of the open quantum system
as
\begin{align}
\rho_s\left(t+dt\right)= & \rho_s\left(t\right)+\sum_{j=0}^{{l-2}}\mathcal{L}_{j}\left(\rho_s\right)dt\nonumber\\
& +\mathcal{O}\left[\sum_{j=0}^{{l-2}}\left(\kappa_j dt\right)^{2}\rho_s\left(t\right)\right].
\label{eq:Master-Kraus}
\end{align}
with Liouvillian $\mathcal{L}_{j}$ describing the incoherent evolution.
Introducing the Kraus operators
\begin{align}
E_{j} & =\sqrt{2\kappa_{j}dt}o_{j},~\mathrm{for}~j\in\left[0,l-2\right]\\
E_{l-1} & =\mathbf{I}-\sum_{j=0}^{l-2}\kappa_{j}dto_{j}^{\dagger}o_{j},
\end{align}
we have an equivalent operation on the system as
\begin{equation}
\rho_s\left(t+dt\right)=\mathcal{E}\left(\rho_s\left(t\right)\right)=\sum_{j=0}^{{l-1}}E_{j}\rho_s\left(t\right)E_{j}^{\dagger}+\mathcal{O}\left[\left(\kappa dt\right)^{2}\rho_s\left(t\right)\right].
\end{equation}
Therefore, the master equation evolution could be approximated by
a repetitive implementation of Kraus operators with $\kappa dt\ll1$.

\subsection{Quantum trajectory simulation}

The evolution of an open quantum system, i.e. the above derived master
equation, could also be interpreted as a quantum jump process: the quantum system changes the environment state from $\left|0\right\rangle $ to $\left|j\right\rangle $ randomly and instantly. Therefore, the dynamics of the system follows a stochastic process
\begin{equation}
\rho_s\left(t+dt\right)\mapsto\frac{E_{j}\rho_s\left(t\right)E_{j}^{\dagger}}{p_{j}},
\end{equation}
with the jump probability of
\begin{equation}
p_{j}=\mathrm{Tr}\left[E_{j}\rho_s\left(t\right)E_{j}^{\dagger}\right].
\end{equation}

In numerical simulations, we can implement the quantum trajectory
simulation by estimating $\left\{ p_{j}\right\} $ and generating
a random number $r\in\left[0,1\right]$, followed by choosing the $j$-th jump
if $\sum_{i=0}^{j-1}p_{i}\leq r<\sum_{i=0}^{j}p_{i}$. In experiments,
the AQuOs applied to the quantum system could be treated as generalized measurements by the ancilla, and the system randomly jumps and evolves conditional on the ancilla outcome $j$. When characterizing
the system states, we repetitively perform the trajectories
and thus obtain the results of unconditioned open system dynamics
by averaging over many different trajectories. As indicated by Eq.~(\ref{eq:Kraus-transformation}),
the choice of the quantum jump operators to realize a quantum trajectory
is not unique, and this property of the quantum trajectory simulation
is known as unravelling~\cite{Wiseman2005}.

\subsection{Quantum measurement\label{subsec:Quantum-measurement}}

For a general measurement process, the ancilla interacts with
the quantum system through a unitary gate that has a similar form as the environment-system interaction. According to Eq.~(\ref{eq:Kraus-Environment}), the output of the ancilla reads
\begin{align}
 & \mathrm{Tr}_{\mathrm{system}}\left\{ U\left[\ket{0}\bra{0}\otimes\rho_{s}\right]U^{\dagger}\right\} \nonumber \\
 & =\Sigma_{j_{1}j_{2}}\ket{j_{1}}\bra{j_{2}}\otimes\mathrm{Tr}\left(E_{j_{1}}\rho_{s}E_{j_{2}}^{\dagger}\right).
\end{align}
Therefore, for the ancilla measurement outcome of $\left|j\right\rangle $,
we have the probability of
\begin{align}
p_{j} & =\left\langle j\right|\mathrm{Tr}_{\mathrm{system}}\left\{ U\left[\ket{0}\bra{0}\otimes\rho_{s}\right]U^{\dagger}\right\} \left|j\right\rangle \nonumber \\
 & =\mathrm{Tr}\left(E_{j}\rho_{s}E_{j}^{\dagger}\right)\nonumber \\
 & =\mathrm{Tr}\left(M_{j}\rho_{s}\right),
\end{align}
with a positive operator valued measure (POVM) $M_{j}=E_{j}^{\dagger}E_{j}$.
Therefore, with the same setup for the AQuO, we can perform arbitrary
POVMs on a qudit by recording the ancilla measurement output. As shown in Fig.~3
in the main text, the circuit to realize AQuOs is equivalent to a
binary-tree structure. Therefore, the realization of POVMs by AQuO
is equivalent to the binary search trees for generalized measurements proposed by Andersson and Oi~\cite{Andersson2008}.

\subsection{Process matrix}

Representing the Kraus operators in a certain basis, the quantum operations
can also be expressed by the process ($\chi$) matrix~\cite{Gilchrist2005,Bhandari2016}
\begin{equation}
\rho_s\rightarrow\sum_{m,n}\chi_{mn}\widetilde{E_{m}}\rho_s\widetilde{E_{n}}^{\dagger}.\label{eq:SI-process}
\end{equation}
Here, for a $d$-dimensional system, the $\chi$ matrix contains $d^{4}-d^{2}$
independent parameters under the constraints  $\sum_{m,n}\chi_{mn}\widetilde{E_{n}}^{\dagger}\widetilde{E_{m}}=\mathbf{I}$.
When characterizing the $\chi$ matrix, we prepare a set of complete
input states for each operation which are linearly independent in the density
matrix representation. Then, we reconstruct $\chi$ matrix by tomography of the output states and represent it with a set of complete orthogonal operators.

For a qubit ($d=2$), we choose Pauli matrices as the operation basis.
For $d=4$, we choose 16 operators that correspond to the tensor product
of two Pauli operators by decomposing the system into two qubits.
For $d=3$, we choose complete orthogonal operators Gell-Mann matrices
$\{\Lambda_{i}\}$:
\begin{align*}
\Lambda_{1}=\begin{pmatrix}1 & 0 & 0\\
0 & 1 & 0\\
0 & 0 & 1
\end{pmatrix}, & \quad\Lambda_{2}=\sqrt{3/2}\begin{pmatrix}0 & 1 & 0\\
1 & 0 & 0\\
0 & 0 & 0
\end{pmatrix},\\
\Lambda_{3}=\sqrt{3/2}\begin{pmatrix}0 & -1 & 0\\
1 & 0 & 0\\
0 & 0 & 0
\end{pmatrix}, & \quad\Lambda_{4}=\sqrt{3/2}\begin{pmatrix}1 & 0 & 0\\
0 & -1 & 0\\
0 & 0 & 0
\end{pmatrix},\\
\Lambda_{5}=\sqrt{3/2}\begin{pmatrix}0 & 0 & 1\\
0 & 0 & 0\\
1 & 0 & 0
\end{pmatrix}, & \quad\Lambda_{6}=\sqrt{3/2}\begin{pmatrix}0 & 0 & -1\\
0 & 0 & 0\\
1 & 0 & 0
\end{pmatrix},\\
\Lambda_{7}=\sqrt{3/2}\begin{pmatrix}0 & 0 & 0\\
0 & 0 & 1\\
0 & 1 & 0
\end{pmatrix}, & \quad\Lambda_{8}=\sqrt{3/2}\begin{pmatrix}0 & 0 & 0\\
0 & 0 & -1\\
0 & 1 & 0
\end{pmatrix},\\
\Lambda_{9}=\sqrt{1/2}\begin{pmatrix}1 & 0 & 0\\
0 & 1 & 0\\
0 & 0 & -2
\end{pmatrix}, & \quad
\end{align*}
where the coefficient in each operator ensures the same
inner product $\mathrm{Tr}(\Lambda_{i}^{\dagger}\Lambda_{i})=3$.

\section{Experimental construction of AQuOs}

For a given target quantum operation, generally in the Kraus operator
representation, we convert it to the system-environment representation
and realize it using the quantum circuit that is shown in Fig.~1 in the main text.

For a rank-2 AQuO, since the ancilla has a dimension of two, we
can directly construct the unitary gate that satisfies
\begin{align}
 & U\left[\ket{g}\bra{g}\otimes\rho_{s}\right]U^{\dagger}\nonumber \\
= & \left|g\right\rangle \left\langle g\right|\otimes E_{0}\rho_{s}E_{0}^{\dagger}+\left|e\right\rangle \left\langle e\right|\otimes E_{1}\rho_{s}E_{1}^{\dagger}\nonumber \\
 & +\left|g\right\rangle \left\langle e\right|\otimes E_{0}\rho_{s}E_{1}^{\dagger}+\left|e\right\rangle \left\langle g\right|\otimes E_{1}\rho_{s}E_{0}^{\dagger}.
\end{align}
As described in the previous sections, the AQuOs require universal control
of the composite system-ancilla. We use the GRAPE method in
our experiments to numerically optimize the drive pulses for the desired
unitary gates with the device parameters that are all experimentally
calibrated. For the optimization, a set of arbitrary and completely independent
states have been chosen, and the control pulses are optimized to achieve the highest average fidelity of the output states compared with the expected ones from the target unitary gates. Since the unitary gates on the qudit with $d=4$ in our experiment only affect the first four Fock levels of the storage cavity mode, four completely independent states which are superpositions of the first four levels of the system in combination with the ground state of the ancilla are enough to constrain the target unitary. More states (over complete) for the constraints lead to a higher precision of the optimized control pulses, but cost more time in the optimization.

For AQuOs with higher ranks, we first decompose the AQuOs into a sequence
of rank-2 AQuOs. All the rank-2 AQuOs could be constructed
following the procedure described above. In the follows, we provide
more details of the schemes presented in the main text.

\subsection{Odd-parity stabilization}

To stabilize the Hilbert space spanned by Fock states of odd parity,
i.e. $\left\{ \left|1\right\rangle ,\,\left|3\right\rangle \right\} $
of the photonic qudit ($d=4$), we have engineered the quantum operation with
two Kraus operators
\begin{equation}
E_{0}=\begin{pmatrix}0 & 0 & 0 & 0\\
0 & 1 & 0 & 0\\
0 & 0 & 0 & 0\\
0 & 0 & 0 & 1
\end{pmatrix},\quad E_{1}=\begin{pmatrix}0 & 0 & 0 & 0\\
1 & 0 & 0 & 0\\
0 & 0 & 0 & 0\\
0 & 0 & 1 & 0
\end{pmatrix}.
\end{equation}
We note that under these operators, any state in the odd-parity subspace
$\mathrm{span}\left\{ \left|1\right\rangle ,\,\left|3\right\rangle \right\} $
is conserved. In contrast, any state outside the odd-parity subspace
will be mapped into the odd-parity subspace.

It is also worth noting that a state satisfying $E_{1}\left|\psi_{D}\right\rangle =0$ would be a dark state that remains invariant under these operators. For instance, $\tilde{E}_{1}=\left|1\right\rangle \left\langle 0\right|+\left|1\right\rangle \left\langle 2\right|$ could map states back to the odd-parity subspace, however, there is a
dark state $\frac{1}{\sqrt{2}}\left(\left|0\right\rangle -\left|2\right\rangle \right)$ outside the odd-parity subspace that could not be stabilized. Our choice of the Kraus operators has a matrix rank of two that avoids the dark states outside
the odd-parity subspace.

\subsection{Two-photon dissipation}

To realize quantum Zeno blockade, we need a two-photon dissipation
operation which converts populations from Fock energy levels $\left|3\right\rangle \rightarrow\left|1\right\rangle $
and $\left|2\right\rangle \rightarrow\left|0\right\rangle $ by eliminating
two photons simultaneously. Conventionally, the two-photon dissipation
has been applied in various experiments to allow the stabilization
of non-classical states~\cite{Touzard2018PRX} and realize
quantum gates between bosonic modes~\cite{Franson2004}. For the continuous
variable case, the master equation of the two-photon dissipation process
is
\begin{equation}
\frac{d}{dt}\rho_s=\frac{\kappa}{2}\left[2a^{2}\rho_s\left(a^{\dagger}\right)^{2}-\left(a^{\dagger}\right)^{2}a^{2}\rho_s-\rho_s\left(a^{\dagger}\right)^{2}a^{2}\right],
\end{equation}
with $\kappa$ being the decay rate. In a truncated space ($d=4$), we
can approximate the process as
\begin{equation}
\frac{d}{dt}\rho_s=\frac{\kappa}{2}\left[2o\rho_s o^{\dagger}-o^{\dagger}o\rho_s-\rho_s o^{\dagger}o\right],
\end{equation}
with $o=\sqrt{2}\left|0\right\rangle \left\langle 2\right|+\sqrt{6}\left|1\right\rangle \left\langle 3\right|$.
For $dt\rightarrow0$, according to Eq.~(\ref{eq:Master-Kraus}),
the equivalent Kraus operator representation is $E_{0}=\mathbf{I}-\frac{\kappa dt}{2}o^{\dagger}o$ and $E_{1}=\sqrt{\kappa dt}o$.
For practical experiments, the AQuOs are implemented with a finite repetitive
duration $t_{\mathrm{int}}$, thus we approximate the two-photon dissipation
process through two Kraus operators
\begin{align}
E_{0} & =\begin{pmatrix}1 & 0 & 0 & 0\\
0 & 1 & 0 & 0\\
0 & 0 & e^{-\kappa t_{{\rm {int}}}} & 0\\
0 & 0 & 0 & e^{-3\kappa t_{{\rm {int}}}}
\end{pmatrix},\\
E_{1} & =\begin{pmatrix}0 & 0 & \sqrt{1-e^{-2\kappa t_{{\rm {int}}}}} & 0\\
0 & 0 & 0 & \sqrt{1-e^{-6\kappa t_{{\rm {int}}}}}\\
0 & 0 & 0 & 0\\
0 & 0 & 0 & 0
\end{pmatrix}.
\end{align}
Here, $\kappa t_{{\rm {int}}}$ is the two-photon dissipation parameter
for each implementation of the operation. For the simulation of quantum
trajectory, the operation is implemented repetitively with an interval
of $t_{{\rm {int}}}$, leading to an effective two-photon dissipation
rate of $\kappa$.

\subsection{AQuOs with arbitrary rank}
In this section we will show a general method to construct an arbitrary
quantum channel with arbitrary ranks of Kraus operators. As proposed by Shen $et~al$.~\cite{Shen2017PRB} and also schematically
illustrated in Fig.~3 in the main text, an AQuO with an arbitrary rank
could be realized by $n$-step adaptive implementations of the elementary rank-2 AQuOs (a binary tree with $n$ layers). Here, we take the full-rank AQuO ($\left\{ E_{k}\right\} $, $k=0,...,15$) for a photonic
qudit with $d=4$ as an example, which is illustrated in Fig.~3 in the main text.

(1) The first step is to decompose the AQuO into Kraus operators at
each layer and construct the binary tree for the target AQuO. The first layer realizes the rank-2 AQuO:
\begin{align}
A_{0}^{\dagger}A_{0} & =E_{0}^{\dagger}E_{0}+...+E_{7}^{\dagger}E_{7},\\
A_{1}^{\dagger}A_{1} & =E_{8}^{\dagger}E_{8}+...+E_{15}^{\dagger}E_{15}.
\end{align}
Therefore, we have
\begin{align}
A_{0} & =\sqrt{E_{0}^{\dagger}E_{0}+...+E_{7}^{\dagger}E_{7}},\\
A_{1} & =\sqrt{E_{8}^{\dagger}E_{8}+...+E_{15}^{\dagger}E_{15}}.
\end{align}
Depending on the ancilla measurement output $\{0,1\}$, corresponding to the subscript of $A$, in the first layer, for example, we have
\begin{equation}
\left(B_{00}A_{0}\right)^{\dagger}\left(B_{00}A_{0}\right)=E_{0}^{\dagger}E_{0}+...+E_{3}^{\dagger}E_{3},
\end{equation}
and thus the second layer realizes
\begin{align}
B_{00} & =\sqrt{E_{0}^{\dagger}E_{0}+...+E_{3}^{\dagger}E_{3}}*A_{0}^{-1},\\
B_{01} & =\sqrt{E_{4}^{\dagger}E_{4}+...+E_{7}^{\dagger}E_{7}}*A_{0}^{-1},\\
B_{10} & =\sqrt{E_{8}^{\dagger}E_{8}+...+E_{11}^{\dagger}E_{11}}*A_{1}^{-1},\\
B_{11} & =\sqrt{E_{12}^{\dagger}E_{12}+...+E_{15}^{\dagger}E_{15}}*A_{1}^{-1}.
\end{align}
Similarly, the third layer realizes
\begin{align}
C_{000} & =\sqrt{E_{0}^{\dagger}E_{0}+E_{1}^{\dagger}E_{1}}*(B_{00}A_{0})^{-1},\\
C_{001} & =\sqrt{E_{2}^{\dagger}E_{2}+E_{3}^{\dagger}E_{3}}*(B_{00}A_{0})^{-1},\\
C_{010} & =\sqrt{E_{4}^{\dagger}E_{4}+E_{5}^{\dagger}E_{5}}*(B_{01}A_{0})^{-1},\\
C_{011} & =\sqrt{E_{6}^{\dagger}E_{6}+E_{7}^{\dagger}E_{7}}*(B_{01}A_{0})^{-1},\\
C_{100} & =\sqrt{E_{8}^{\dagger}E_{8}+E_{9}^{\dagger}E_{9}}*(B_{10}A_{1})^{-1},\\
C_{101} & =\sqrt{E_{10}^{\dagger}E_{10}+E_{11}^{\dagger}E_{11}}*(B_{10}A_{1})^{-1},\\
C_{110} & =\sqrt{E_{12}^{\dagger}E_{12}+E_{13}^{\dagger}E_{13}}*(B_{11}A_{1})^{-1},\\
C_{111} & =\sqrt{E_{14}^{\dagger}E_{14}+E_{15}^{\dagger}E_{15}}*(B_{11}A_{1})^{-1},
\end{align}
and the last layer realizes
\begin{equation}
\begin{split}D_{0000}=E_{0}*(C_{000}B_{00}A_{0})^{-1},\\
...\quad\\
D_{1111}=E_{15}*(C_{111}B_{11}A_{1})^{-1}.
\end{split}
\end{equation}

Therefore, after the implementation of four layers of AQuOs, we
realize the Kraus operators $E_{k}=D_{ghij}C_{ghi}B_{gh}A_{g}$ with $k=8^{g}+4^{h}+2^{i}+1^{j}$, as expected. Here $g,h,i,j \in \{0,1\}$ are the measurment outcomes at each layer, respectively. In practice, the operators
may not be invertible, which induces difficulties in calculating the
target Kraus operators for the $m$-th layer $(m>1)$. For example, if
$C_{ghi}B_{gh}A_{g}$ is not invertible, then $D_{ghij}$ in not well defined. To solve this problem, we can replace $(C_{ghi}B_{gh}A_{g})^{-1}$
with $(C_{ghi}B_{gh}A_{g}+\epsilon \mathbf{I}_{d\times d})^{-1}$, where $\epsilon\ll1$ is a small positive number that makes $(C_{ghi}B_{gh}A_{g}+\epsilon \mathbf{I}_{d\times d})^{-1}$ meaningful. Finally, $(C_{ghi}B_{gh}A_{g}+\epsilon \mathbf{I}_{d\times d})^{-1}$ will be cancelled by $(C_{ghi}B_{gh}A_{g})$, so the constructed AQuOs will be unaffected for an infinitesimal $\epsilon \mathbf{I}_{d\times d}$.

(2) The second step is to construct the unitary gates for the rank-2
AQuOs at each layer in the binary-tree construction for the target AQuO (Fig.~3 in the main text). In the example of the rank-16 AQuO, when the operators $\left\{ A_{g}\right\} $,
$\left\{ B_{gh}\right\} $, $\left\{ C_{ghi}\right\} $, $\left\{ D_{ghij}\right\} $
are calculated, we can construct the unitary gates on the composite system-ancilla according to Eq.~(\ref{eq:Kraus-Environment}). Here, the unitary gates constructed by $\left\{ B_{gh}\right\} $, $\left\{ C_{ghi}\right\} $,
$\left\{ D_{ghij}\right\} $ may not be unitary in the whole space of the photonic qudit. For example,
\begin{equation}
U_{D\_ghi}=\begin{pmatrix}D_{ghi0} & *\\
D_{ghi1} & *
\end{pmatrix}\quad
\end{equation}
may not be a unitary gate because $D_{ghi0}^{\dagger}D_{ghi0}+D_{ghi1}^{\dagger}D_{ghi1}$
may not be $\mathbf{I}_{4\times4}$ for the photonic qudit with $d=4$,
while $U_{D\_ghi}$ could be a unitary gate for a subspace of the qudit.
We can use an unnormalized state $C_{ghi}B_{gh}A_{g}\ket{\Psi}$
($\ket{\Psi}$ is an arbitrary state in the whole space of the qudit) to
represent the arbitrary state in the subspace.

\begin{figure*}[hbt]
\includegraphics{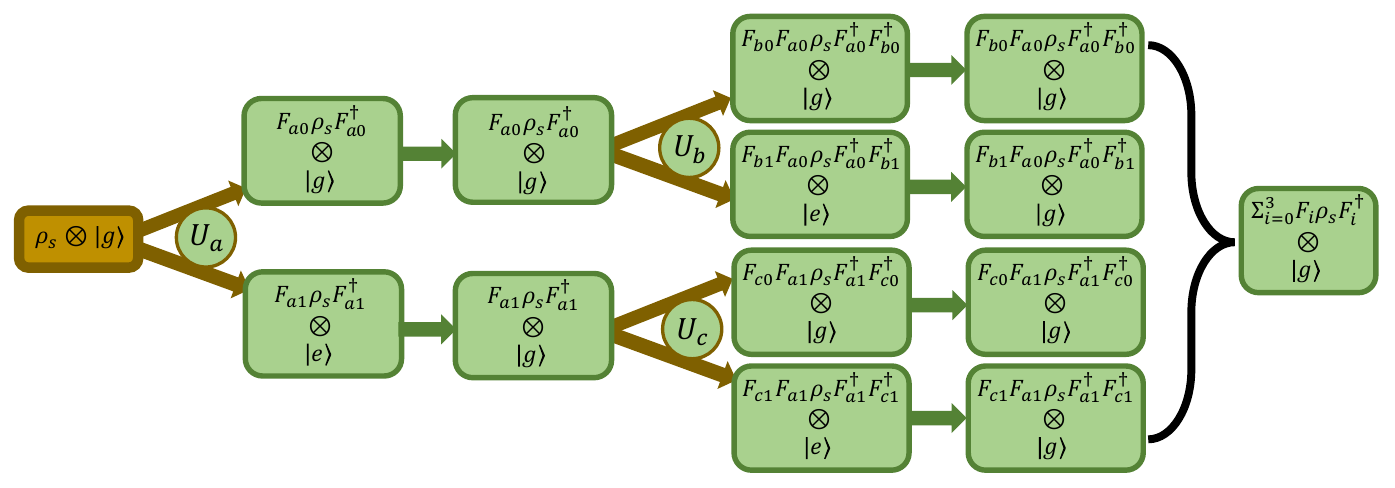} \caption{\textbf{Flowchart illustration of the realization of the strictly incoherent operation (SIO).}}
\label{fig:S_SIO}
\end{figure*}

\subsection{Construction of Two SIOs}

The rank-2 strictly incoherent operation (SIO) demonstrated in the main text is given by two Kraus operators
\begin{equation}
E_{0}=\begin{pmatrix}1 & 0 & 0 & 0\\
0 & 1 & 0 & 0\\
0 & 0 & 0 & 0\\
0 & 0 & 0 & 0
\end{pmatrix},\quad E_{1}=\begin{pmatrix}0 & 0 & 0 & 0\\
0 & 0 & 0 & 0\\
0 & 0 & 1 & 0\\
0 & 0 & 0 & 1
\end{pmatrix}.\quad\label{eq:1stSIO}
\end{equation}
This SIO could be realized directly via a rank-2 AQuO.

The rank-4 SIO is given by four Kraus operators
\begin{equation}
\begin{split}F_{0}=\begin{pmatrix}1 & 0 & 0 & 0\\
0 & 0 & 0 & 0\\
0 & 0 & 0 & 0\\
0 & 0 & 0 & 0
\end{pmatrix},\quad F_{1}=\begin{pmatrix}0 & 0 & 0 & 0\\
0 & 1 & 0 & 0\\
0 & 0 & 0 & 0\\
0 & 0 & 0 & 0
\end{pmatrix},\quad\\
F_{2}=\begin{pmatrix}0 & 0 & 0 & 0\\
0 & 0 & 0 & 0\\
0 & 0 & 1 & 0\\
0 & 0 & 0 & 0
\end{pmatrix},\quad F_{3}=\begin{pmatrix}0 & 0 & 0 & 0\\
0 & 0 & 0 & 0\\
0 & 0 & 0 & 0\\
0 & 0 & 0 & 1
\end{pmatrix}.\quad
\end{split}
\label{eq:2ndSIO}
\end{equation}
According to the procedure in the previous section, the rank-4 SIO could
be realized adaptively in a two-layer binary-tree construction, as shown by
the flowchart in Fig.~\ref{fig:S_SIO}. As an example of the high-rank
AQuOs, we provide the details of the operations as follows. For a binary-tree with two layers, there are three unitary gates on the composite system-ancilla:
\begin{equation}
U_{a}=\begin{pmatrix}F_{a0} & *\\
F_{a1} & *
\end{pmatrix},\quad
\end{equation}
\begin{equation}
U_{b}=\begin{pmatrix}F_{b0} & *\\
F_{b1} & *
\end{pmatrix},\quad
\end{equation}
\begin{equation}
U_{c}=\begin{pmatrix}F_{c0} & *\\
F_{c1} & *
\end{pmatrix}.\quad
\end{equation}
Each unitary gate together with a measurement and reset of the ancilla realizes a rank-2 AQuO and the corresponding Kraus operators
(\{$F_{a0},F_{a1}$\}; \{$F_{b0},F_{b1}$\}; \{$F_{c0}$, $F_{c1}$\})
in Fig.~\ref{fig:S_SIO} are given by
\begin{equation}
F_{a0}=\begin{pmatrix}1 & 0 & 0 & 0\\
0 & 1 & 0 & 0\\
0 & 0 & 0 & 0\\
0 & 0 & 0 & 0
\end{pmatrix},\quad F_{a1}=\begin{pmatrix}0 & 0 & 0 & 0\\
0 & 0 & 0 & 0\\
0 & 0 & 1 & 0\\
0 & 0 & 0 & 1
\end{pmatrix},\quad
\end{equation}
\begin{equation}
F_{b0}=\begin{pmatrix}1 & 0 & 0 & 0\\
0 & 0 & 0 & 0\\
0 & 0 & 0 & 0\\
0 & 0 & 0 & 0
\end{pmatrix},\quad F_{b1}=\begin{pmatrix}0 & 0 & 0 & 0\\
0 & 1 & 0 & 0\\
0 & 0 & 0 & 0\\
0 & 0 & 0 & 0
\end{pmatrix},\quad
\end{equation}
\begin{equation}
F_{c0}=\begin{pmatrix}0 & 0 & 0 & 0\\
0 & 0 & 0 & 0\\
0 & 0 & 1 & 0\\
0 & 0 & 0 & 0
\end{pmatrix},\quad F_{c1}=\begin{pmatrix}0 & 0 & 0 & 0\\
0 & 0 & 0 & 0\\
0 & 0 & 0 & 0\\
0 & 0 & 0 & 1
\end{pmatrix}.
\end{equation}
It is easy to verify that $F_{0}=F_{b0}F_{a0}$, $F_{1}=F_{b1}F_{a0}$,
$F_{2}=F_{c0}F_{a1}$, $F_{3}=F_{c1}F_{a1}$, and $F_{0},\,F_{1},\,F_{2},\,F_{3}$
are the set of Kraus operators [Eq.~(\ref{eq:2ndSIO})] that represent the target rank-4 SIO
[$\varepsilon_{F}(\rho_{s})=\Sigma_{i=0}^{3}F_{i}\rho_{s}F_{i}^{\dagger}$]
for the photonic qudit. It is worth noting that the construction of this
SIO is equivalent to a projection measurement of the qudit onto
the Fock state basis, i.e. the first AQuO implements the measurement
to distinguish the subspace $\mathrm{span}\left\{ \left|0\right\rangle ,\left|1\right\rangle \right\} $
and $\mathrm{span}\left\{ \left|2\right\rangle ,\left|3\right\rangle \right\}$, while the subsequent AQuOs $\left\{ F_{b0},F_{b1}\right\} $ and $\left\{ F_{c0},F_{c1}\right\}$ realize the projections on the Fock states.

\begin{figure*}[hbt]
\includegraphics{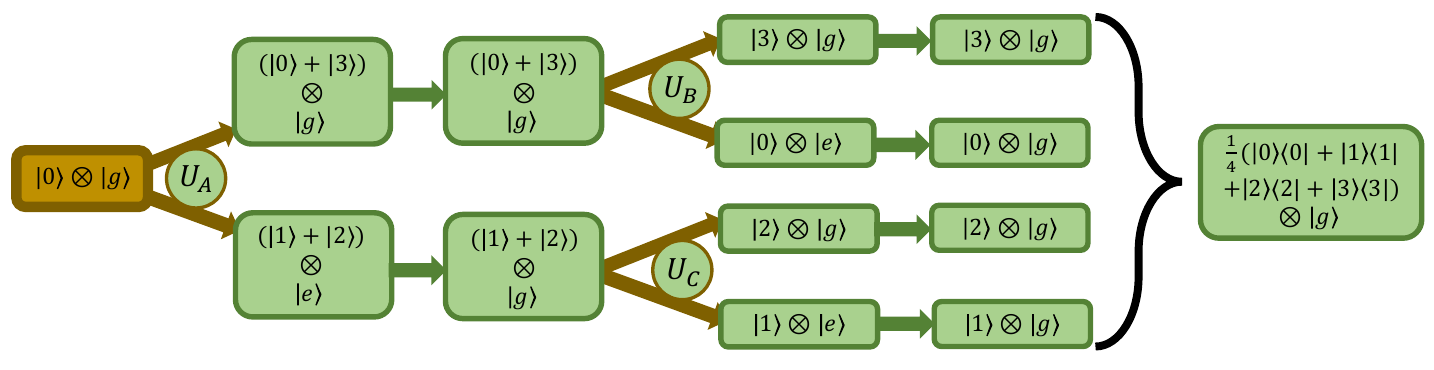} \caption{\textbf{Flowchart illustration of the preparation of the maximally mixed state of the qudit by the AQuO architecture.} The composite system-ancilla is initialized to $\left|0\right\rangle \otimes\left|g\right\rangle $.}
\label{fig:S_State-preparation}
\end{figure*}

\begin{figure}[hbt]
\includegraphics{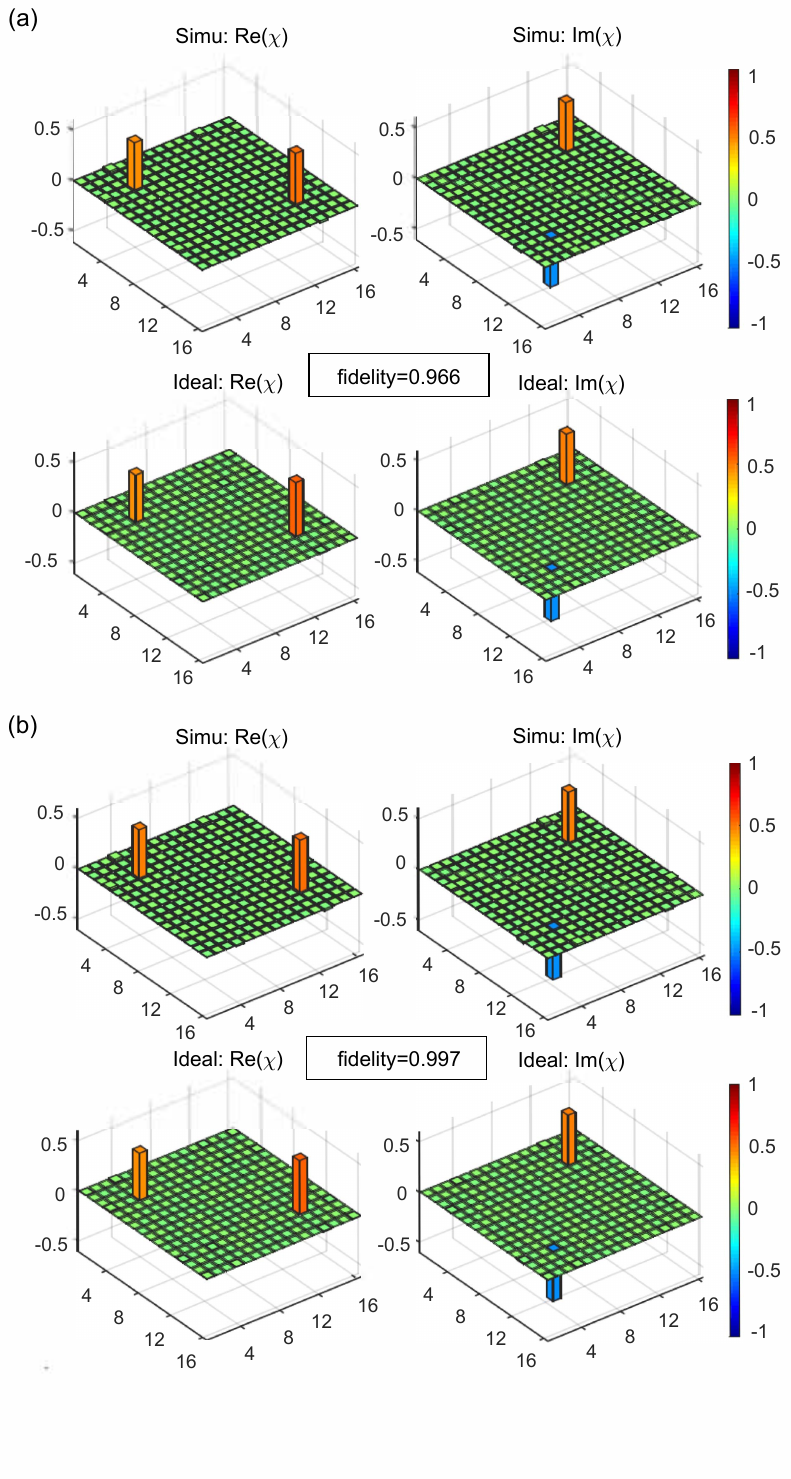} \\
\caption{\textbf{Simulated process fidelity of the odd-parity stabilization operation with (a) and without (b) the ancilla decoherence.} Real (left) and imaginary (right) parts of $\chi$ matrix are respectively shown. $\chi_ {\mathrm{simu}}$ (top panels) is numerically simulated through QuTip with the experimental pulses and system parameters, while $\chi_{\mathrm{ideal}}$ (bottom panels) is calculated by ideal numerical matrices. The numbers in the $x$ and $y$ axes correspond to the 16 operators which have the same Pauli matrix forms for two qubits but different basis states $\{\ket{g}\otimes(\ket{1},\ket{3}),\ket{e}\otimes(\ket{1},\ket{3})\}$. The color bars represent the value of each matrix element.}
\label{Fig-gate-fidelity}
\end{figure}

\subsection{State preparations}

In the main text, to show the capability of AQuOs in arbitrary quantum
state preparation, we have prepared two typical states, i.e. the maximum
coherent state (MCS, $\left|\Psi\right\rangle =\frac{1}{\sqrt{d}}\sum\left|d\right\rangle $)
and the maximally mixed state (MMS, $\rho_s=\frac{1}{d}\sum_{j=0}^{d}\left|d\right\rangle \left\langle d\right|\equiv\frac{1}{d}\mathbf{I}_{d\times d}$).
For the pure state preparation, it could be realized by the previously
demonstrated arbitrary unitary gate, by either GRAPE or
SNAP~\cite{Krastanov2015} approach, that transforms the initially pure state Fock
$\ket{0}$ of the photonic qudit to an MCS $\ket{\Psi}=\frac{1}{2}(\ket{0}+\ket{1}+\ket{2}+\ket{3})$.
However, the MMS could not be prepared via unitary gates.

Figure~\ref{fig:S_State-preparation} shows a flowchart of the scheme
for preparing the MMS. The system is firstly initialized to a pure
state of $\ket{0}\otimes\ket{g}$ (brown box). Circles represent unitary
gates for the rank-2 AQuOs. At the first layer, $U_{A}$ maps the
initial state to a pure state $\frac{1}{2}(\ket{0}+\ket{3})\otimes\ket{g}+\frac{1}{2}(\ket{1}+\ket{2})\otimes\ket{e}$.
Then, the entanglement between the ancilla and the qudit is erased by flipping the ancilla state $\left|e\right\rangle $ to $\left|g\right\rangle $ while keeping $\left|g\right\rangle $ unchanged. After this step, the system is prepared in a mixed state. Then, in the second layer $U_{B}$ or $U_{C}$ is applied adaptively according to the
measurement output of the ancilla in the first layer. Here, $U_{B}$ maps state $\frac{1}{\sqrt{2}}(\ket{0}+\ket{3})\otimes\ket{g}$
to state $\frac{1}{\sqrt{2}}\ket{3}\otimes\ket{g}+\frac{1}{\sqrt{2}}\ket{0}\otimes\ket{e}$
and $U_{C}$ maps state $\frac{1}{\sqrt{2}}(\ket{1}+\ket{2})\otimes\ket{g}$ to state
$\frac{1}{\sqrt{2}}\ket{2}\otimes\ket{g}+\frac{1}{\sqrt{2}}\ket{1}\otimes\ket{e}$.
By resetting the ancilla to $\left|g\right\rangle $, the final state
of the composite system becomes  $\rho=\frac{1}{4}\mathbf{I}_{4\times4}\otimes\left|g\right\rangle \left\langle g\right|$.

\begin{figure*}[hbt]
\includegraphics[width=0.8\textwidth]{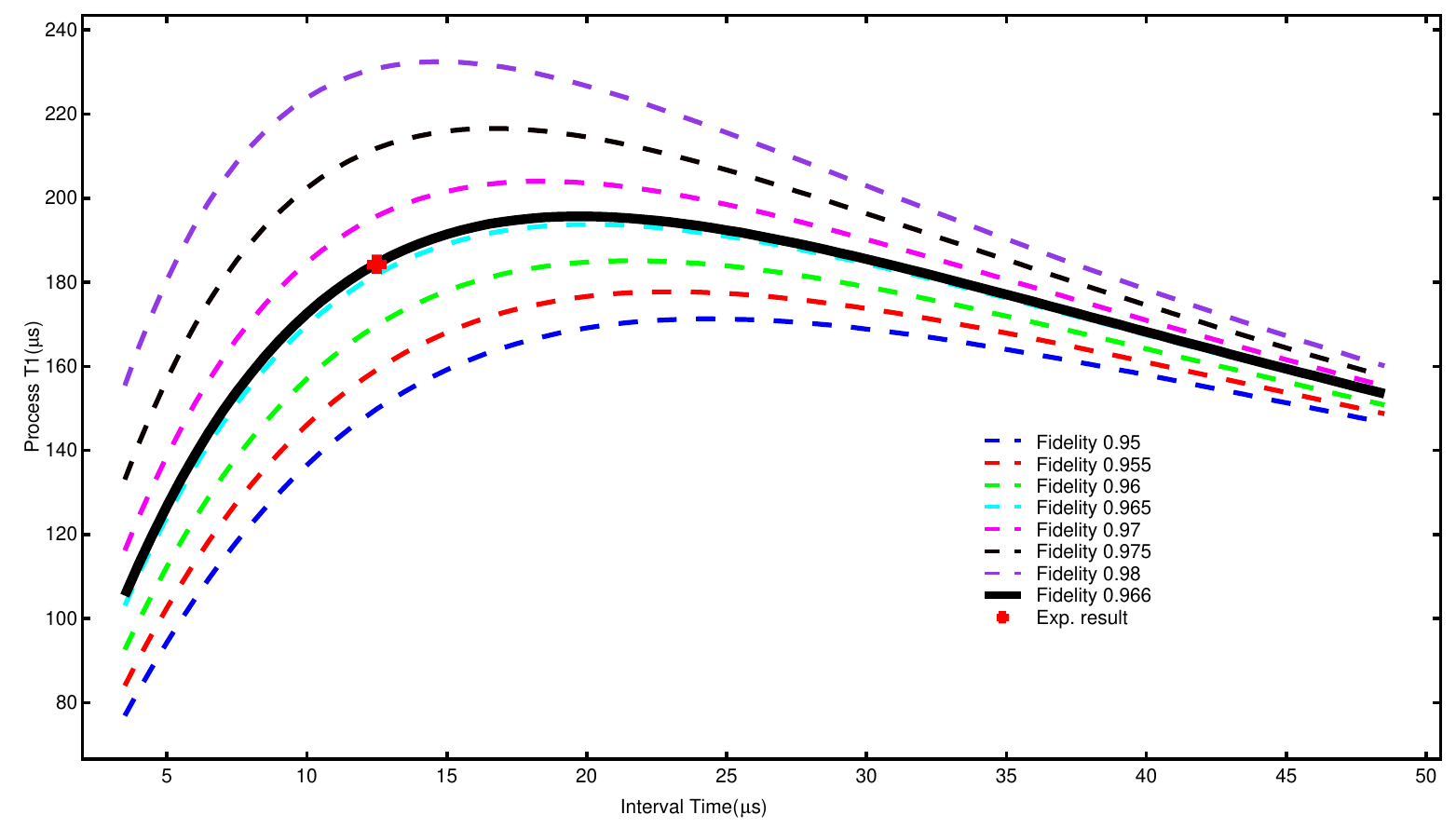} \caption{\textbf{Process fidelity decay time $T_1$ of the stabilized odd-parity subspace as a function of the interval time for different gate fidelities.}}
\label{Figs-process-T1}
\end{figure*}

\begin{figure*}[tb]
\includegraphics{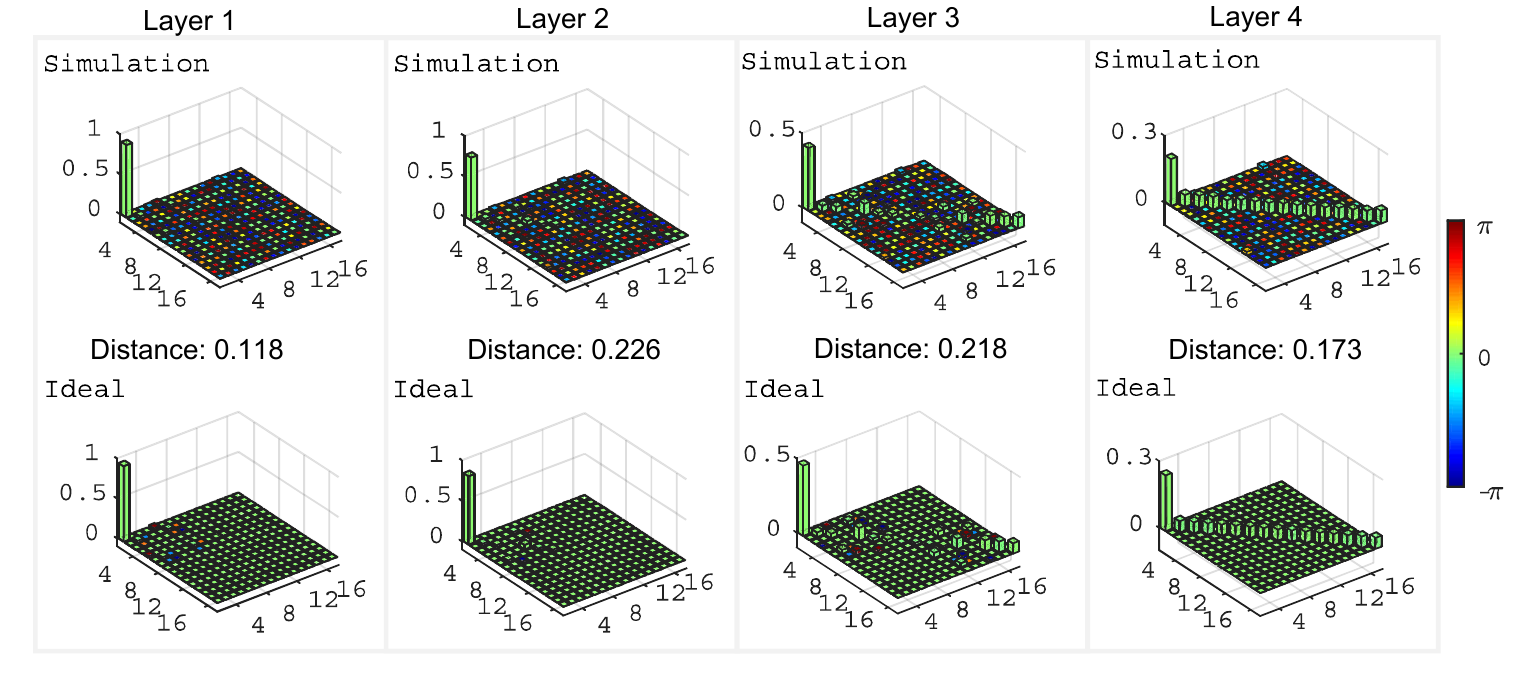} \caption{\textbf{$\chi$ matrix of the 4-level SIC-POVMs at each layer of the binary-tree method.} Top panels: $\chi_{\mathrm{simu}}$ is numerically simulated through QuTip with the experimental
GRAPE pulses and system parameters including the decoherence. Bottom panels: $\chi_{\mathrm{ideal}}$ is obtained from the numerically simulated ideal AQuO. The height of each matrix bar indicates the absolute value of corresponding element in the $\chi$ matrix, and the color corresponds to the complex angle. Diamond distance between $\chi_{\mathrm{simu}}$ and $\chi_{\mathrm{ideal}}$ at each layer of the binary-tree construction is also shown.}
\label{Fig_s_ChiMatrix_SICPOVM}
\end{figure*}

\begin{figure}[tb]
\includegraphics{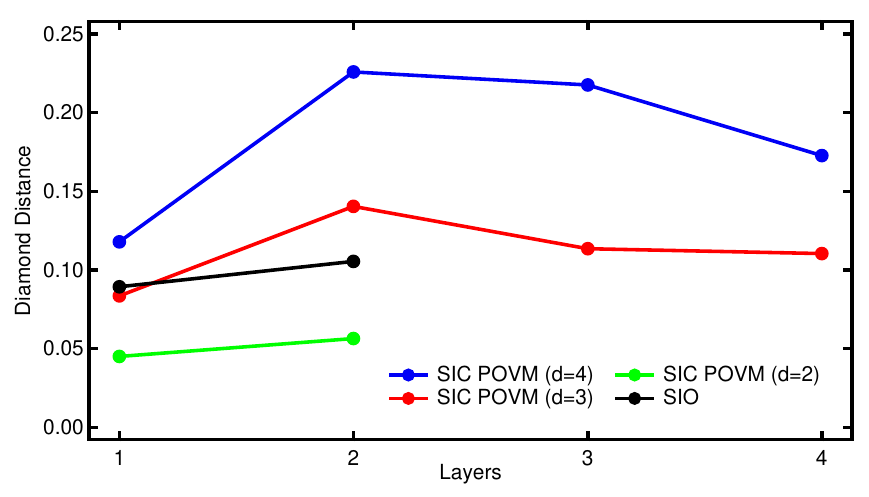} \caption{\textbf{Diamond distance at each layer of the binary-tree construction for different AQuOs.}}

\label{Fig_S_DiamondDistance}
\end{figure}

\begin{figure}[tb]
\includegraphics{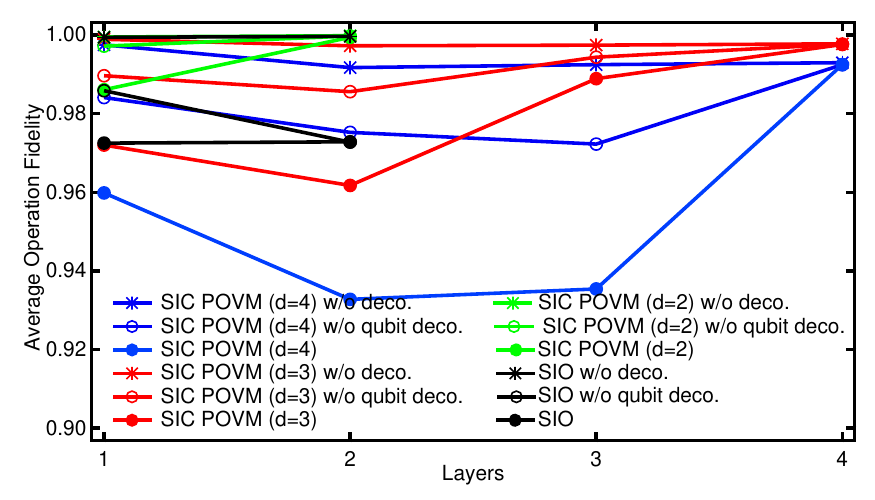} \caption{\textbf{Average operation fidelity at each layer of the binary-tree construction for different AQuOs.} Solid dots, hollow dots, and stars respectively correspond to the practical experimental condition including all decoherence and imperfections, the condition excluding the ancilla decoherence, and the condition excluding all decoherence. Different colors represent different AQuOs.}
\label{Fig_S_average_state_fidelity}
\end{figure}

\section{Imperfections}

\subsection{Optimal interval time for quantum trajectory simulations}

In the quantum trajectory simulations, we choose a certain interval and repetitively implement the AQuOs. In principle, the continuous evolution of an open quantum system could be simulated perfectly with
ideal unitary operations on the composite system-ancilla with infinitesimal
operation time and fast adaptive control. However, in practical experimental system, there is non-negligible decoherence of the ancilla and the photonic qudit. Because of the finite $T_{1}$ and $T_{2}$ of the ancilla, the fidelities
of the unitary gates are limited for the GRAPE pulses that are optimized based on an ideal ancilla. Considering the errors of the unitary gates, if we implement AQuOs too frequently with a short interval between sequential AQuOs, significant
gate errors would accumulate in the experiments. On the other hand, the
qudit would bear more photon loss errors due to energy dissipation
if the interval is too long. The trade-off between gate errors and qudit photon loss errors leads to an optimal interval time in the simulation of quantum trajectories. In this section, we systematically investigate the optimal interval via numerical simulations of the experimental procedure including the system decoherence.

Figure~\ref{Fig-gate-fidelity} shows the simulated fidelities of the unitary
gates in the odd-parity stabilization operation with
and without the ancilla decoherence in QuTip~\cite{Johansson2012,Johansson2013}.
The process ($\chi$) matrix in Fig.~\ref{Fig-gate-fidelity} are obtained
based on the initial states $\ket{g}\otimes(\ket{0},\ket{1},\ket{2},\ket{3})$ and measured in four bases $\{\ket{g}\otimes(\ket{1},\ket{3}),\ket{e}\otimes(\ket{1},\ket{3})\}$. However, in the presence of decoherence and the imperfection of the GRAPE pulse, the final photonic qudit could have non-negligible leakage to the higher energy levels after the unitary gate, which cannot be detected in the simulation. Therefore, the trace of the obtained output density matrix is smaller than 1. To overcome this problem, we treat the leakage as an appropriate maximally mixed state within the final detection space such that the trace of the output density matrix becomes 1. This treatment of the leakage outside the qudit computational space is valid and similar to the real experiment where the leaked qudit states will be decoded to a mixed state of the ancilla for the final detection. From the numerical results, we conclude that the gate error is mainly attributed to the decoherence of the ancilla qubit, and the resulted gate (process) fidelity is about $96.6\%$.

For the practical implementation of the unitary gates on the composite
system-ancilla, a random error occurring during the gate evolution would
induce unpredictable and random output due to the complex GRAPE control
pulses. Therefore, we could approximate such a decoherence process
as a depolarization channel acting on the system with an equal probability for all possible physical errors during the unitary gate. For a $d$-level
system, the depolarization channel is given by
\begin{align}
\mathcal{E}_{\mathrm{dpl}}(\rho) & =p\frac{\mathbf{I}_{d\times d}}{d}+\left(1-p\right)\rho,
\end{align}
where $\mathbf{I}_{d\times d}$ is the $d\times d$ identity matrix
and $p$ is the depolarization probability. This indicates that the output
is equivalent to replace the quantum state by an MMS with a probability $p$.

Then, the process fidelity of the operation compared with the ideal identity
operation is
\begin{align}
F_{\chi} & =\left(1-p\right)\frac{d^{2}-1}{d^{2}}+\frac{1}{d^{2}}
\end{align}
For our case of $d=4$, the probability of the depolarization channel is given by
\begin{align}
p & =1-(0.966-\frac{1}{16})/\frac{15}{16}=0.036.
\end{align}

In our experiments, we actually realize an effective operation approximated
by $\mathcal{\mathcal{E}_{\mathrm{dpl}}}\circ\mathcal{E}_{\mathrm{Target}}$
when we are targeting to realize an AQuO $\mathcal{E}_{\mathrm{Target}}$.
For the quantum trajectory simulations, we repetitively implement
the AQuOs with an interval of $t_{\mathrm{int}}$. With $p\ll1$, the whole
dynamics of the system follows the target quantum trajectory evolution,
while the system simultaneously couples to a depolarization channel
with a rate of $p/t_{\mathrm{int}}$ and a single photon decay channel
with a rate of $\kappa_{\mathrm{s}}$, where $\kappa_{\mathrm{s}}$
is the intrinsic decay rate of the storage cavity. So, for the quantum
trajectory simulations, there is an optimal interval time that balances the photon decay rate, the dephasing rate $\propto K\kappa_{s}t_{\mathrm{int}}$ induced
by the photon decay error and self-Kerr effect ($K$), and the effective
depolarization rate $p/t_{\mathrm{int}}$.

To optimize $t_{\mathrm{int}}$, we numerically simulate the process
fidelity decay time $T_{1}$ with the above depolarization approximation. Figure~\ref{Figs-process-T1}
shows the process fidelity decay time $T_{1}$ as a function of $t_{\mathrm{int}}$ for different gate fidelities. The black solid line correspondes to a gate fidelity of $96.6\%$ and the red cross correspondes to the experimental result for $t_{\mathrm{int}}=12.5~\mathrm{\mu s}$.
The simulation results suggest the process fidelity decay time $T_{1}$ could exceed $180~\mathrm{\mu s}$ if the optimal interval time around $15~\mathrm{\mu s}$ is chosen and a gate fidelity $\geq96.6\%$ is achieved. In practice, there
is also the Kerr effect that would induce extra phases and should be compensated
by carefully designing the GRAPE pulses. Therefore, the GRAPE pulse should
be numerically optimized for each given interval time. Considering
the finite numerical precision and parameter fluctuations in the numerical
optimization, there is a fluctuation of the infidelity varying from case
to case. Comparing the numerical estimation and the experimental result, the experimental process fidelity decay time $T_{1}=184.3~\mathrm{\mu s}$ agrees well with simulation.

\subsection{Process matrix and diamond distance}

To characterize the performance of the AQuOs in practical quantum systems, we compare the practical and the ideal $\chi$ matrices of a rank-16 AQuO for a symmetric informationally complete positive operator-valued measure (SIC-POVM) realized in Fig.~3 in the main text. The $\chi$ matrices are obtained after implementing each layer of the binary tree. $\chi_{\mathrm{simu}}$ is numerically simulated through QuTip with the experimental GRAPE pulses and system parameters including the decoherence, while $\chi_{\mathrm{ideal}}$ is obtained from the numerically simulated ideal AQuO. The results are shown in Fig.~\ref{Fig_s_ChiMatrix_SICPOVM}. It is worth mentioning that after implementing the entire four layers of a SIC-POVM we could reconstruct the input quantum state by recording the probability of measuring each POVM element.

In this section, to characterize the performance of the SIC-POVM we treat it as a quantum operation which could be verified by the $\chi$ matrix. Since there are numerical errors in the ideal $\chi$ matrix due to finite numerical precisions, we set the matrix element's complex angle to be 0 if the element's absolute value is smaller than $10^{-6}$ for the clarity of the ideal $\chi$ matrix. Comparing the simulated and the ideal matrices, it can be seen that our experimental system could realize the AQuO with small imperfections at each step of the binary-tree implementation.

To quantify the performance of each quantum operation in practice,
we evaluate the diamond distance ($D_{\diamond}$) between the numerically simulated ideal AQuO and the physically realizable one. For simplicity, the physically realizable AQuOs are also numerically simulated with the exact experimental pulses and parameters. Then we have
\begin{align}
D_{\diamond} & =||\chi_{\mathrm{simu}}-\chi_{\mathrm{ideal}}||_{\diamond}\\
 & =\mathrm{sup}_{\rho}||(\chi_{\mathrm{simu}}\otimes\mathbf{I})(\rho)-(\chi_{\mathrm{ideal}}\otimes\mathbf{I})(\rho)||_{1},
\end{align}
where the trace norm is defined as
\begin{equation}
||T||_{1}\equiv\mathrm{Tr}\sqrt{T^{\dagger}T}.
\end{equation}
The $D_{\diamond}$ quantifies the difference between two AQuOs as
it is related to the upper bound of the success probability
\begin{equation}
p_{\mathrm{succ}}=\frac{1}{2}+\frac{1}{4}D_{\diamond}
\end{equation}
in the single-shot discrimination of two quantum operations by using an ancilla in the discrimination ($0\leq D_{\diamond}\leq2$).
When $D_{\diamond}=2$, the two operations could be perfectly distinguished.

In Fig.~\ref{Fig_s_ChiMatrix_SICPOVM}, the calculated diamond distance
between the practically simulated and ideal $\chi$ matrices are labeled. To further study the contribution of the imperfections when realizing AQuOs, $D_{\diamond}$'s of SIC-POVMs for $d=2$, 3 and 4 
are also investigated, and the results are summarized in Fig.~\ref{Fig_S_DiamondDistance}.
From the results, we find $D_{\diamond}$ for most operations increases with the number of layers in the binary-tree construction. There are also some extraordinary results: $D_{\diamond}$'s of SIC-POVMs for $d=4$ (blue points) and $d=3$ (red points) decrease when implementing the entire layers of the binary tree. Although there are coherent errors from the pulse imperfections and numerical calculations, we conjecture that these coherent errors will be reduced by the incoherent operations. Based on the process matrix of implementing the SIC-POVM ($\chi$ matrix of entire layers) in Fig.~\ref{Fig_s_ChiMatrix_SICPOVM}, the process of SIC-POVMs corresponds to a depolarization channel which is an incoherent operation. That is the reason for the diamond distance to decrease when implementing entire layers of the binary tree. In Fig.~\ref{Fig_S_DiamondDistance}, two SIOs (black points) are presented: the implementation of one layer corresponds to the rank-2 SIO with the Kraus operators in Eq.~(\ref{eq:1stSIO}) and implementation of two layers corresponds to the rank-4 SIO with the Kraus operators in Eq.~(\ref{eq:2ndSIO}). For comparison, we also show the average operation fidelities
\begin{equation}
\mathcal{F}(\chi_{\mathrm{ideal}},\chi_{\mathrm{simu}})=\left[\mathrm{Tr}\sqrt{\sqrt{\chi_{\mathrm{ideal}}}\chi_{\mathrm{simu}}\sqrt{\chi_{\mathrm{ideal}}}}\right]^{2}
\end{equation}
of different AQuOs in Fig~\ref{Fig_S_average_state_fidelity}.

\section{State reconstruction and SIC-POVMs}

\subsection{Density matrix reconstruction}

As shown in the main text and also Sec.~\ref{subsec:Quantum-measurement},
the POVM could be realized by recording the outputs of the ancilla with
the AQuO architecture. In contrast to many other experiments of POVM,
the probabilities of POVM elements acting on state $\rho_{s}$ could
all be measured unconditionally by the AQuO setup. Therefore, it is convenient to reconstruct the density matrix of a quantum system
$\rho_s$ from measurement results of a SIC-POVM, since $\rho_s$ can be easily
reconstructed by a linear transformation of the probabilities $p_{k}=\mathrm{Tr}(M_{k}\rho_{s})$ from the SIC-POVM $\left\{ M_{k}\right\} $. For the schematic shown in Fig.~3(b) in the main text, the system's SIC-POVM has 16 elements for a qudit with $d=4$. In the experiment, $p_{k}$ directly equals to $p_{ghij}$, the joint probability of the four measurement outcomes of the ancilla in the binary-tree construction. For example, $k=7$ corresponds
to $ghij=0111$.

In experiments, a non-positive semi-definite density matrix might be reconstructed from $\left\{ p_{k}\right\} $ by the linear transformation due to physical errors. So, maximum likelihood estimation instead of a linear transformation is used in practice to find a set of probabilities $\left\{ p_{k}^{\prime}\right\} $, based on which the reconstructed density matrix is Hermitian and positive semi-definite and has the minimum distance from the one based on $\left\{ p_{k}\right\}$.

Subjecting the reconstructed density matrix to Hermitian and semi-definite,
we evaluate the quantum state fidelity by
\begin{equation}
\mathcal{F}(\rho_{\mathrm{T}},\rho)=\left[\mathrm{Tr}\sqrt{\sqrt{\rho_{\mathrm{T}}}\rho\sqrt{\rho_{\mathrm{T}}}}\right]^{2}.
\end{equation}
Here, $\rho$ is the density matrix derived in experiment and $\rho_{\mathrm{T}}$
is the ideal target density matrix. The relative entropy of coherence
is given by
\begin{equation}
\mathcal{C}(\rho)=S(\rho_{\mathrm{diag}})-S(\rho).
\end{equation}
Here, $S$ is the von Neumann entropy and $\rho_{\mathrm{diag}}$
is the state obtained by removing the off-diagonal elements of $\rho$
in the Fock basis~\cite{Streltsov2017}.

\subsection{Calibration of SIC-POVMs}

The elements of a SIC-POVM read
\begin{equation}
M_{k}=\frac{1}{d}\ket{\phi_{k}}\bra{\phi_{k}},
\end{equation}
where $k\in\{0,1,...,d^{2}\}$ and the basis state $\left|\phi_{i}\right\rangle $
can be generated by operating $d^{2}$ displacement operators on the fiducial
vectors~\cite{Bent2015PRX}, i.e.
\begin{equation}
\left|\phi_{i}\right\rangle =D_{jk}\left|\phi_{f}\right\rangle ,
\end{equation}
where $j,k=\{1,2,...,d\}$, $d$ is dimension of system, $\left|\phi_{f}\right\rangle$ is the fiducial state, and $D_{jk}$ is the displacement operator that is given by~\cite{Bent2015PRX,Renes2004SICPOVM}
\begin{equation}
D_{jk}=\omega_{d}^{jk/2}\Sigma_{m}\omega_{d}^{jm}\ket{k\oplus m}\bra{m}.
\end{equation}
Here, $\omega_{d}=e^{i\frac{2\pi}{d}}$, $\left|m\right\rangle $
is Fock state, and $\bigoplus$ represents the addition modulo $d$. The
fiducial vectors (not unique) in our experiment for dimensions $d=2$, 3, 4
are
\begin{equation}
\left|\phi_{f}\right\rangle _{d=2}=\begin{pmatrix}0.8881+0.0000i\\
0.3251-0.3250i
\end{pmatrix},\quad
\end{equation}
\begin{equation}
\left|\phi_{f}\right\rangle _{d=3}=\begin{pmatrix}0.8124+0.0000i\\
0.1677+0.2911i\\
0.2386+0.4125i
\end{pmatrix},\quad
\end{equation}
and
\begin{equation}
\left|\phi_{f}\right\rangle _{d=4}=\begin{pmatrix}0.2012+0.0000i\\
-0.3076+0.2570i\\
0.0000+0.4857i\\
-0.1064+0.7427i
\end{pmatrix},\quad
\end{equation}
respectively.

\begin{figure}[hbt]
\includegraphics{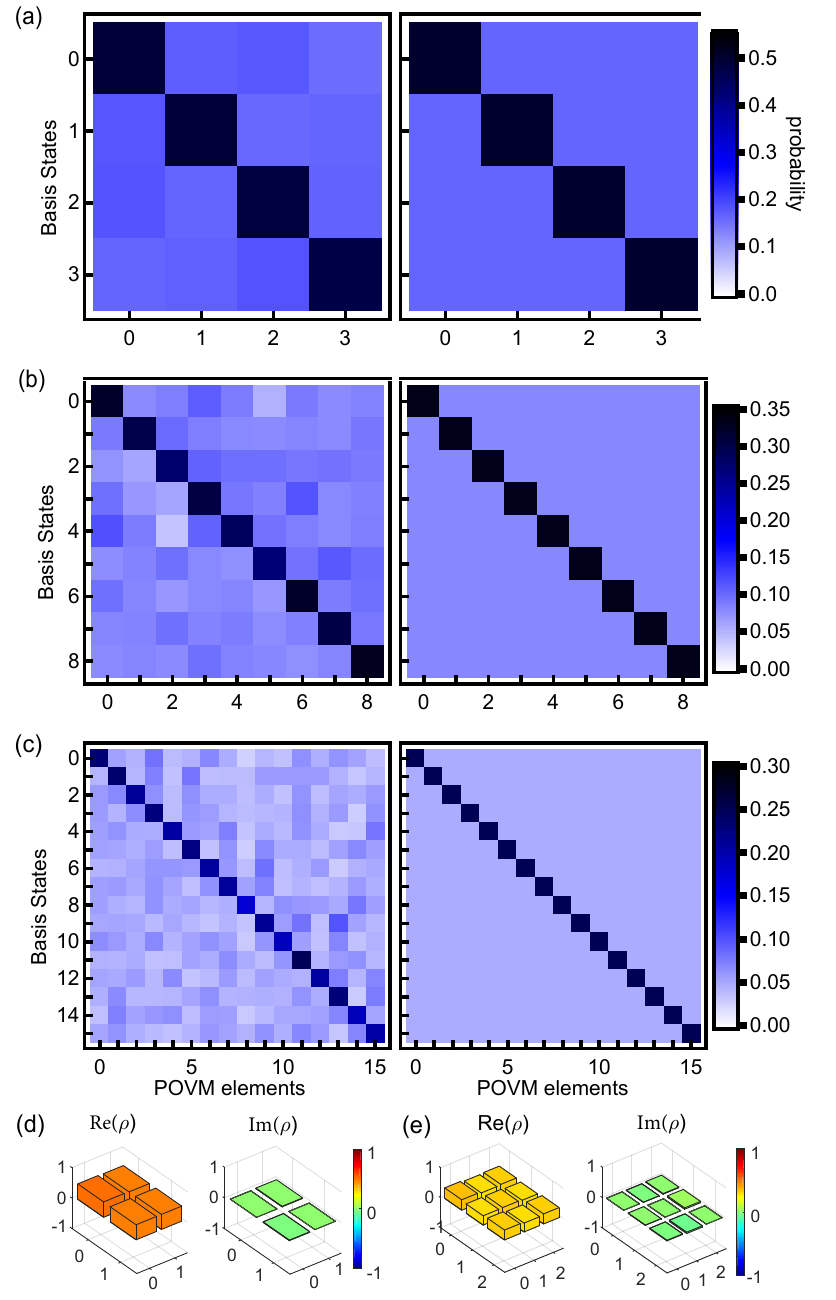} \caption{\textbf{Characterization of SIC-POVMs}. \textbf{(a-c)} Characterization of SIC-POVMs for $d=2$, 3, 4: experiment (left) and theory (right). $x$-axis presents different SIC-POVM elements $M_{x}=1/d\ket{\phi_{x}}\bra{\phi_{x}}$, $y$-axis indicates different prepared basis states $\ket{\phi_{y}}$, and the color represents the probabilities of prepared states measured along the SIC-POVM elements ($p\left(x,y\right)=\mathrm{Tr}\left[M_{x}\left|\phi_{y}\right\rangle \left\langle \phi_{y}\right|\right]$). \textbf{(d, e)} State tomography of the maximum coherent state for $d=2$ (fidelity is 99.44$\%$) and $d=3 $ (fidelity is 95.59$\%$), respectively, by using SIC-POVMs.}
\label{fig:S_SICPOVM}
\end{figure}

To evaluate the performance of a SIC-POVM, we first prepare all
basis states, and then characterize these states via the SIC-POVM.
Figure~\ref{fig:S_SICPOVM} shows the experimental results of the state
tomography of all basis states by using the SIC-POVM. The average fidelities for all basis states of the SIC-POVM are $96.80\%$, $89.61\%$, and $83.63\%$ for $d=2$, $3$, and $4$, respectively. We notice that the fidelities of some states are higher than the average fidelity of the basis states, and we suppose it is because the probability of measuring the basis state is mostly distributed along only one SIC-POVM element ($25\%$ for $d=4$) while the general states are not the case. For example, the most probability of the measurement outcomes for the MCS ($d=4$) is evenly distributed along four different SIC-POVM elements, while for the MMS the measurement probability is evenly distributed along all SIC-POVM elements. We suppose that a state with more evenly distributed probability of measuring the SIC-POVM elements will be robust against errors in the state reconstruction and tends to have a higher state fidelity.

\subsection{Comparison with Wigner tomography}

Conventionally, the states encoded in the storage cavity are characterized
by Wigner tomography in the superconducting quantum circuit platform.
In this work, we demonstrate the quantum state reconstruction by a
SIC-POVM unconditionally, and find that our AQuO approach can greatly
save measurement time compared to the conventional approach. Therefore, SIC-POVMs provide an efficient means of quantum state tomography for a quantum system with any finite dimensions~\cite{Bent2015PRX,Renes2004SICPOVM}.

Comparing the SIC-POVM and conventional Wigner tomography, the required experimental measurement time of quantum state tomography on a qudit with $d=4$ by a SIC-POVM (about $3$ minutes) is about $200$ times shorter than that by Wigner tomography (about $10$ hours with the same experimental setup). However, the states reconstructed
by the SIC-POVM have slightly lower fidelities ($92.52\%$ for the MCS and $98.83\%$ for the MMS) than those by Wigner tomography ($99.57\%$ for the MCS and $99.73\%$ for the MMS). This is because that we use a four-layer binary-tree method to implement the SIC-POVM for the prepared states, and the operation errors accumulate in this process (four steps of rank-2 AQuOs for $d=4$).


%